\documentclass[oldversion]{aa}
\pdfoutput=1

\usepackage{graphicx}
\usepackage{txfonts}
\usepackage{natbib}
\usepackage{subfigure}
\usepackage{color}
\usepackage{xspace}
\usepackage{soul}
\newcommand{\diff}[1]{\frac{d}{d#1}}

\newcommand{\cstar}{$c_\star$\xspace}

\newcommand{\sfr}{star formation rate\xspace}

\newcommand{\mfe}{$\rm{[Fe/H]}$\xspace}

\newcommand{\feh}{$\rm{[Fe/H]}$\xspace}

\begin{document}


   \title{The Dynamical and Chemical Evolution of Dwarf Spheroidal Galaxies with \texttt{GEAR}}

   \author{Y. Revaz\inst{1} 
          \and 
	  P. Jablonka\inst{1,2}
}

  \institute{Laboratoire d'Astrophysique, \'Ecole Polytechnique F\'ed\'erale de Lausanne (EPFL), 1290 Sauverny, Switzerland\\
        \and
        GEPI, Observatoire de Paris, CNRS UMR 8111, Universit\'e Paris Diderot,  F-92125, Meudon, Cedex, France \\
}	

   \date{Received -- -- 20--/ Accepted -- -- 20--}

 
   \abstract{We present a fully parallel chemo-dynamical Tree/SPH
     code, \texttt{GEAR}, which allows to perform high resolution
     simulations with detailed chemical diagnostics. Starting from the
     public version of \texttt{Gadget-2}, we included the complex
     treatment of the baryon physics: gas cooling, star formation law,
     chemical evolution and supernovae feedback.  We qualified the
     performances of \texttt{GEAR} with the case of dSph galaxies.
     \texttt{GEAR} conserves the total energy budget of the systems to better
     than $5\%$ over $14\,\rm{Gyr}$ and proved excellent convergence
     of the results with numerical resolution.  We showed that models
     of dSphs in a static Euclidean space, where the expansion of the
     universe is neglected are valid.  In addition, we tackled some of
     the existing open questions in the field, like the stellar mass
     fraction of dSphs and its link with the predicted dark matter
     halo mass function, the effect of the supernova feedback, the
     spatial distribution of the stellar populations, and the origin
     of the diversity in star formation histories and chemical
     abundance patterns.  Strong supernovae driven winds seem
     incompatible with the observed metallicities and luminosities.
     Despite the fact that newly formed stars are preferentially found
     in the galaxy central parts, turbulent motions in the gas can
     quickly erase any metallicity gradient.  The variety in dSph
     properties result from a range of total masses as well as from a
     dispersion in central densities. The latter is also seen in the
     haloes emerging from a $\Lambda$CDM cosmogony.
 }


   \maketitle

%

\section{Introduction}\label{introduction}

The first motivation of the present work was the clear need for an
extension of \citet{revaz09}, which sought to identify the dominant
physical processes at the origin of the dynamical and chemical
properties of dwarf spheroidal galaxies (dSphs).  While allowing a
critical step forward with a large set of simulations, the excellence
of the \texttt{treeAsph} code used in this work was somewhat lessen by
its serial numerical structure \citep{serna96,alimi03}, hampering the
possibility of high resolution simulations due to large computing
time.

The analysis of other Tree/SPH codes, that include chemical evolution
schemes and were applied to the formation and evolution of dwarf
galaxies, demonstrates that the specific astrophysical aims of each
study influence if not drive the numerical implementations.  Using
colour-magnitude diagrams as observational probes, \citet{carraro01}
modeled their gas particles as closed box models and used the
instantaneous recycling approximation.
\citet{marcolini06,marcolini08} focused on the response of the ISM to
the supernovae explosion and introduced an arbitrary number of star
formation episodes. \citet{kawata06} following \citet{kawata03},
combined kinetic and thermal feedbacks, which could generate galactic
winds in elliptical galaxies.  \citet{okamoto10} investigated the
  effect of different recipes of kinetic feedback driving galactic
  winds on the global properties of the Milky Way satellite model
  galaxies.  \citet{read06} looked for a mass boundary distinguishing
between essentially pure dark haloes and small stellar systems. They
chose a unique global yield reproducing the metallicity of the
inter-galactic medium at z=3. \citet{stinson07} and
\citet{governato10} applied the \citet{stinson06}'s blast wave recipe
for the feedback. \citet{sawala10} used the multiphase scheme for the
interstellar medium developed by \citet{scannapieco05,scannapieco06},
and constrained the fraction of metals given to the cold gas phase,
later defining the final metallicity for a given stellar mass, on the
metallicity-relation of dwarf galaxies.  \citet{valcke08} and later
\citet{schroyen11} took into account stellar winds and supernovae
explosions but only consider the global metallicity $Z$.  
  \citet{ricotti05} stopped their simulations very early on (z$\sim$8)
  but included continuum radiative transfer to compare the global
  properties of dSphs with a simulated sample of galaxies formed
  before reionization.

We aimed at a new N-body code, hereafter \texttt{GEAR}, which fulfills
a number of requirements among which primarily the capability to offer
high spatial resolution together with detailed chemical diagnostics,
and to follow the galaxy evolution over a full Hubble time, either in
isolation or in a cosmological context. Because our approach to
constrain the scenarios of galaxy evolution involves the confrontation
with observed stellar metallicity distributions and stellar abundance
ratios, we seriously evaluated the impact of the supernovae feedback
on these quantities, and further probed all ingredients in the
parametrization of the star formation.  
While it would be illusive to expect tracing in detail the star
formation processes at the scale of molecular clouds, still we tried
to release as much as possible ad hoc assumptions, and based our
modeling on physical grounds. 
Great care was devoted to trace all sources and sinks of energy and guaranty 
an accurate budget of gains and losses. We also controlled the convergence 
of the results.

\texttt{GEAR} is a flexible fully parallel chemo-dynamical Tree/SPH
code, which we applied to tackle some of the questions left open after
\citet{revaz09}, such as the gas motions and the final radial
distribution of the stellar population in dwarf spheroidal galaxies.
It also allowed to expand on a number of new subjects, in particular
the question of the relation between galaxies in isolation and their
parent haloes in cosmological simulations, and clarify the final
baryonic fraction of galaxies as a function of their halo masses.

The paper is organized as follows: the code and the implementation of
physical processes are described in Sect.~\ref{code_gear}. The initial conditions
are detailed in Sect.~\ref{initial_conditions}. The robustness of the code is discussed in
Sect.~\ref{robustness}. The role and impact of the main parameters are evaluated in
Sect.~\ref{parameter_effect}, while Sect.~\ref{mass_profile} focuses on the initial mass and central gas
density. Sect.~\ref{regimes_of_starformation} describes the different regimes of star formation
for low mass galaxies and makes the first detailed comparison with the
observations. Sect.~\ref{gradients} addresses the question of the spatial
distribution of the stellar population as a function of their age and
metallicity. Sect.~\ref{stellar_mass_fraction} investigates the link between stellar masses in
galaxies and their parent dark matter haloes.  Sect.~\ref{conclusion} summarizes
our results.

%

\section{\texttt{GEAR}}\label{code_gear}

We supplemented the public version of \texttt{Gadget-2}  \citep{springel05} with the complex treatment of the baryon physics:
gas cooling, star formation law, chemical evolution and supernovae feedback.

\subsection{The gas cooling}

The interstellar medium (ISM) is modeled as an ideal, inviscid gas
with an adiabatic index $\gamma=5/3$.  The continuity equation is
inserted into the first law of thermodynamics in order to follow the
evolution of the gas specific internal energy. The variation of the
internal energy depends on the mechanical forces, the artificial
viscosity and the radiative cooling expressed through a cooling
function $\Lambda(\rho,T)$.  Following the integration scheme of
\texttt{Gadget-2}, we use the entropy function $A$ (see \citet{springel02} for
the exact definition) instead of the internal energy as independent
thermodynamic variable.  The variation of $A$ is simplified as a sum
of two terms corresponding to the contributions of the artificial
viscosity \citep{springel05} and the cooling: %
	\begin{equation}
        \diff{t} A = \diff{t}A \bigg)_{\rm{visc}}   + \diff{t}A \bigg)_{\rm{cool}},
	\label{dadt}
	\end{equation}
where:
	\begin{equation}
	\diff{t}A \bigg)_{\rm{cool}} =  -(\gamma-1)\,\rho^{-\gamma}\,\Lambda(\rho,T).
	\end{equation}

For temperatures hotter than $10^4\,\rm{K}$, the cooling function is
calculated following the metallicity dependent prescription of
\citet{sutherland93}. Below $10^4\,\rm{K}$, the cooling of H$_2$ and
HD molecules are taken into account as well as the atoms of oxygen,
carbon, silicon and iron \citep{maio07}, see \citet{revaz09} for more
details.

In dense ($\rho > 0.01\,\rm{m_{H}/cm^3}$) and warm-hot regions
($T>10^4\,\rm{K}$), the cooling time may be much shorter than the
dynamical time requiring extremely short time steps. In such a case,
we use the isochoric approximation (see for example,
\citet{springel01}), assuming a constant density $\rho$. The
integration of Eq.~\ref{dadt} is performed using adaptive time steps
set to a fraction of 2\% of the cooling time defined as~: %
    \begin{equation}
    t_{\rm{cool}} = \frac{A}{  {\frac{d}{dt}}A\big)_{\rm{cool}}  }.
    \end{equation}
We do not take into account the heating of the gas by a cosmic background UV field.  
Its effect was  investigated in a number of former studies
(e.g, \citet{sawala10,okamoto10}), because it could contribute to
evaporate the gas in small haloes ($v_{\rm{max}}\le 12\,\rm{km/s}$).

  However, the ISM is known to be strongly non homogeneous down to
  very small scales, reaching densities $\sim 10^6\,\rm{m_{H}/cm^3}$
  \citep[e.g.,][and references therein]{omont07,beuther07}, for
  which the cooling time is very short.  As long as those
  over-densities are not properly resolved, including an additional
  heating source, as the UV background is not a real gain, because
  the cooling will be underestimated due to the lack of resolution.
  This is without taking into account that the gas may be
  self-shielded strongly complicating the effect of the UV field.
  In this picture, improving the cooling by following precisely all 
  cooling agents involved in is also unprofitable.
As no heating process is taken into account here, 
the integration of Eq.~\ref{dadt} does not suffer from numerical instabilities 
and an implicit integration scheme is not required.

\subsection{Star formation law}

The complex and poorly understood formation processes is implemented
using the phenomenological prescription proposed by \citet{katz92} and
\citet{katz96}.

A gas particle  becomes eligible to star formation when it fulfills the following physical conditions:
{\it i})   the particle is collapsing (its velocity divergence is negative),
{\it ii})  its density is higher than a threshold  $\rho_{\rm{sfr}}$,
{\it iii}) its temperature is lower than a threshold $T_{\rm{sfr}}$.

In a time interval $\Delta t$, an eligible gas particle of mass $m_{\rm{g}}$ has a probability $p_{\star}$ to form a stellar particle
of mass $m_{\star}$ (\citet{springel03}):
        \begin{equation}
        p_{\star} = \frac{m_{\rm{g}}}{m_{\star}}\left[ 1-\exp\left(  -\frac{c_\star}{t_{\rm{g}}}\Delta t  \right) \right],
        \label{pstar}
        \end{equation}
where $c_\star$ is the star formation parameter and $t_{\rm{g}}$ the local free fall time.
This ensures a star formation law independent of both the time step $\Delta t$ and $m_{\star}$:
        \begin{equation}
        \frac{d \rho_\star}{dt} = \frac{c_\star}{t_{\rm{g}}}\rho_{\rm{g}},
        \label{sfr}
        \end{equation}
in which  $\rho_{\rm{g}}$ is the gas density.

Each gas particle can form a number of stellar particles, this number
is set by the parameter $N_\star$.  Each new stellar particle
represents an ensemble of stars sampling an initial mass function
(IMF) with different slopes, $\alpha$, in four stellar mass ranges
\citep{kroupa01}, taking into account  the systematic effects due to unresolved binaries :
     \begin{equation}
        \alpha=\left\{ 
        \begin{array}{lcr}
        -0.3, &\rm{if}&                      $m$ <0.08\,\rm{M_\odot} \\
        -1.8, &\rm{if}& 0.08\,\rm{M_\odot} < $m$ <0.50\,\rm{M_\odot} \\
        -2.7, &\rm{if}& 0.50\,\rm{M_\odot} < $m$ <1.00\,\rm{M_\odot} \\
        -2.3, &\rm{if}& $m$ > 1.00\,\rm{M_\odot} \\
        \end{array}
        \right.
        \label{kroupa}
        \end{equation}

        The minimal and maximal stellar masses, that are considered in
        the IMF, are $0.05$ and $50\,\rm{M_\odot}$, respectively.  The
        stellar particles receive the positions, velocities, and
        chemical abundances of their parent gas particles.  These gas
        and stars are dynamically decoupled, therefore their evolution
        in the phase space are quickly diverging.

        In \texttt{GEAR}, the stellar particles are numerically
        considered as an independent class of particles with their own
        structure. This is necessary in order to optimize the memory
        needed to store their chemical properties.  \texttt{GEAR}
        keeps the possibility to order all types of particles
        following the Peano-Hilbert curve similarly to
        \texttt{Gadget-2}.  This allows to take advantage of the fast
        cache-memory available on modern processors.

We set $T_{\rm{sfr}}$ to $3 \times 10^4\,\rm{K}$. The exact value of $T_{\rm{sfr}}$ is not crucial.
It is chosen here to avoid the formation of stars in hot gas.
The other parameters, $c_\star$, $\rho_{\rm{sfr}}$, and the maximal number of stellar particles
that may be created from one gas particle $N_\star=m_{\rm{g}}/m_{\star}$ are discussed in detail 
in Section~\ref{parameter_effect}.

\subsection{Chemical evolution}\label{stellar_ejecta}

The galactic evolution is influenced by two main features of the
stellar evolution: on the one hand the nucleosynthesis, on the other
hand, the stellar feedback (winds or supernova explosions).  In the
following, we only consider the effect of supernovae.  Indeed,
stellar winds from intermediate mass stars inject little power to the
ISM as compared to the explosion of supernovae \citep{leitherer92}.
Moreover, the evolution of the galactic systems that we are considering
is already well constrained by the abundances of $\alpha$-elements
(magnesium) and iron and their ratios. Both types of elements are
produced by Type Ia (SNeIa) and of Type II (SNeII) supernovae.

For a given stellar particle, the number of stars ending their
lifetime is computed at each dynamical time step. These stellar
lifetime depends on the metallicity at the creation of the particle
and are taken from \citet[private communication]{kodama97}.  For an
accurate resolution, we impose a maximum time step of $0.2\,\rm{Myr}$.
This corresponds to less than 10\% of the smallest lifetime of the
exploding stars in our simulation (about $3\,\rm{Myr}$).  The stellar
mass dependent yields of SNeII for stars between $8$ and
$50\,\rm{M_{\odot}}$ are taken from \citet{tsujimoto95}.  We adopt the
model of \citet{kobayashi00} for the explosion of SNeIa.  The
progenitors of SNeIa have main-sequence masses between $3$ and
$8\,\rm{M_\odot}$, and evolve into C+O white dwarfs (WDs). These white
dwarfs can form two different types of binary systems, either with
main sequence stars or with red giants (see \citet{revaz09} for more
the details on the adopted parameters). The nucleosynthesis products
of SNeIa are taken from the model W7 of \citet{iwamoto99}.  Contrary
to \citet{kobayashi00} we do not prevent the SNeIa explosion at
[Fe/H] below $-1$.  We note that the nucleosynthesis tables and the
IMF can be changed easily in \texttt{GEAR} .

The stellar chemical ejecta are distributed among
the nearest neighbors following the SPH scheme
\citep[e.g.,][]{wiersma09,revaz09}.  First, the gas density $\rho_i$ and
corresponding smoothing length $h_i$ are determined at the location of
each stellar particle $i$.  Both quantities are computed similarly to
the case of the gas particles \citep{springel05}. The mass of
metals ejected from particle $i$ and attributed to particle $j$ is
computed through the weights $w_{ij}$:
        \begin{equation}
        w_{ij} = \frac{m_j W(r_{ij},h_i)}{\rho_i},
        \label{wij}
        \end{equation}

where $W$ is the SPH kernel function, $m_j$ is the mass of the
particle $j$ and $r_{ij}$ the distance between the particles $i$ and
$j$.  By default, the number of neighbors used is set to
$N_{\rm{ngb}}=50$. The effect of varying $N_{\rm{ngb}}$ is discussed
in Section~\ref{nngb}.  In order to improve the conservation of 
energy during the process of mass ejection, the velocities and the 
entropy of the particles are modified as described in Appendix~\ref{appendix1}.

The chemical abundances are calculated with respect to the solar
abundances of \citet{anders89}.  The V-band luminosities are derived
following the stellar population synthesis model of \citet{vazdekis96}
computed for the  revised \citep{kroupa01}'s IMF.  Where necessary, the
luminosities are inter- and extra- polated in age an metallicity
using a bi variate spline.

\subsection{Feedback}\label{feedback}

Each explosion of supernova injects an energy $\epsilon_{\rm{SN}}
E_{\rm{SN}}$ into the ISM, with $E_{\rm{SN}}=10^{51}\,\rm{ergs}$ and
$\epsilon_{\rm{SN}}$ the feedback efficiency factor, which is further
discussed in Section~\ref{esn}.

The supernova explosion feedback is definitively a complex mechanism,
which led to a large number of reports discussing both its numerical
implementation and its physical operation modes.  One can broadly
define three different types of methods discussed in the literature:
{\it i}) The so-called thermal feedback in which the energy is
released through heating of the gas : either in an homogeneous ISM
\citep{gerritsen97,mori97,thacker00,sommerlarsen03,brook04,stinson06}
or in a multiphase ISM
\citep{yepes97,hultman99,marri03,scannapieco06}.
{\it ii}) A kinetic feedback in which energy is mechanically released
\citep{navarro93,springel03,dallavecchia08}.

We were very careful in ensuring  {\it an accurate budget of
gains and losses of energy},
even in presence of a strong feedback.  We also checked that the
energy from the supernovae, as predicted by stellar evolution, was
fully injected in the galactic system.  This particular point is not
always granted, especially when using a stochastic implementation of
the kinetic feedback \citep{springel03,dallavecchia08}. In practice,
at each time step $[t,t+\Delta t]$, we calculate the amount of
energy do be released during the explosions of SNe. This quantity
$E_{\rm{SN}}(t)$ is the sum of the contributions of SNeIa and SNeII:
        \begin{equation}
        E_{\rm{SN}}(t) = \sum_i \Delta E_{i,\rm{SN}} = \sum_i m_{i,\star,0}\,\,\left[ n_{i,\rm{II}}(t)\,E_{\rm{II}} + n_{i,\rm{Ia}}(t)\,E_{\rm{Ia}}  \right],
        \end{equation}
where $m_{i,\star,0}$ is the initial mass of the stellar particle $i$,
and $n_{i,\rm{II}}(t)$ and $n_{i,\rm{Ia}}(t)$ are the corresponding
numbers of supernovae SNeII and SNeIa per unit mass during $\Delta
t$.  The integral of $E_{\rm{SN}}(t)$ over time is then compared to
the energy injected in the system during the feedback procedure.

In the case of a kinetic feedback, the particles are pushed away,
forming a wind with a velocity of the order of $100\,\rm{km/s}$
\citep{springel03,dallavecchia08}. This hardly conserves energy to a
satisfactory level.  Energy budget is improved if the wind
particles are decoupled from the other gas particles or if the
dynamical time steps are shorten before the advent of the wind.
Unfortunately, this latter solution increases significantly the
computation time.  Moreover, the choice of a given wind speed limits
the amount of energy deposited in each neighboring particle. Hence the
total feedback energy derived from stellar evolution is not always
fully transferred, especially for slow winds. Consequently, this
technique implicitly imposes strong winds.

For all these reasons, we chose a thermal feedback.  At each time
step, the energy $E_{\rm{SN}}$ is distributed in
thermal form among the surrounding particles using the same weight
$w_{ij}$ as for the distribution of the chemical
elements (Eq.~\ref{wij}). When combined with radiative cooling, such a
thermal feedback can be inefficient.  Indeed, the radiative cooling of
the heated gas becomes very large and the injected energy is
instantaneously lost \citep{katz92}.  In \texttt{GEAR}, the cooling of
the gas particle receiving the thermal feedback is switched off for a
short adiabatic period of time, $t_{\rm{ad}}$ (a few Myr)
\citep{gerritsen97,mori97,thacker00,sommerlarsen03,brook04,stinson06}.
As both supernovae types release the same amount of energy,
the switch is applied to gas particles receiving feedback from both
SNeII and SNeIa, without distinction. Thus, the feedback from SNIa may 
be more effective than the one used by other authors \citep[for example]{stinson06}.
The impact of $t_{\rm{ad}}$ is discussed in Section~\ref{tad}.  We
carefully checked that our implementation of feedback combined with
the cooling led to converging results with decreasing time step and
increasing resolution.

We did not use any time step limiter to prevent possible
numerical problems induced by a pre-shock timestep too long compared
to the shock timescale. This time step limiter is important when
large differences in temperature occur between the pre- and post-
shock media \citep{saitoh09,merlin10}.  It is not critical in our
case, as the net heating of the ISM due to SNe explosions never
exceeds a factor $20$.


\section{Initial conditions}\label{initial_conditions}

\subsection{Evolution of small dark matter haloes}\label{dmh}

\begin{figure}
  \resizebox{\hsize}{!}{\includegraphics[angle=0]{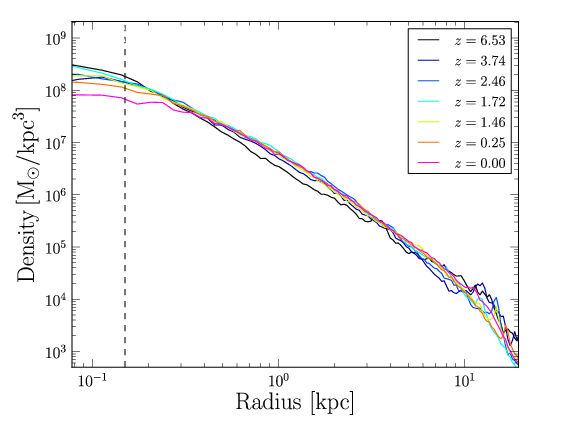}}
  \caption{Evolution of the dark halo density profile for a halo   mass of $6\times 10^8\,\rm{M_\odot}$ as a function of redshift. The dashed
line indicates the limit of the resolution corresponding to the gravitational softening.}
  \label{fig1_condinit}
\end{figure}

We ran a $\Lambda$CDM cosmological simulation in order to study the
profiles of dark haloes with masses between $10^8$ and
$10^9\,\rm{M_\odot}$, that are typical of dSphs
\citep{walker07,battaglia08}. The volume of the simulation is $2^3\,h^{-3}\,\rm{Mpc^3}$ and contains
$134'217'728$ dark matter particles which results in a particle mass
of $4.6\times 10^3\rm{M_\odot/h}$ and a softening length of $150\,\rm{pc/h}$.
We took the cosmological parameters from the concordance $\Lambda$CDM
flat universe based on WMAP V data combined with the
Baryon Acoustic Oscillations (BAO) in the distribution of galaxies and distances measurements from 
Type Ia supernovae (SN) observations \citep{hinshaw09,komatsu09}~:
$\Omega_{m}=0.279$, $\Omega_{\Lambda}=0.721$ and $h_{0}=0.7$.
We extracted the
dark haloes at different redshifts using the HOP algorithm \citep{eisenstein98}.
The halo mass is defined as the mass inside $r_{\rm{200}}$, the radius
at which the density of matter is 200 times the critical density of
the Universe;  the halo density profiles have a NFW shape \citep{navarro96,navarro97}.
We found $144$ haloes with masses in the range from $10^8$ to $10^9\,\rm{M_\odot}$.
The analysis of these haloes showed two main features:

(i) for $50\%$ of the haloes, the density profiles in \emph{physical} coordinates (as opposed to \emph{comoving} coordinates)
are in place  at a redshift of $6$, i.e., at about $1\,\rm{Gyr}$ (see Fig.\ref{fig1_condinit}).
Whilst the Universe was in average about $350$ times denser at $z=6$ than at $z=0$ (in physical coordinates), 
the density profiles of the dark haloes are similar. 

\begin{figure}
  \resizebox{\hsize}{!}{\includegraphics[angle=0]{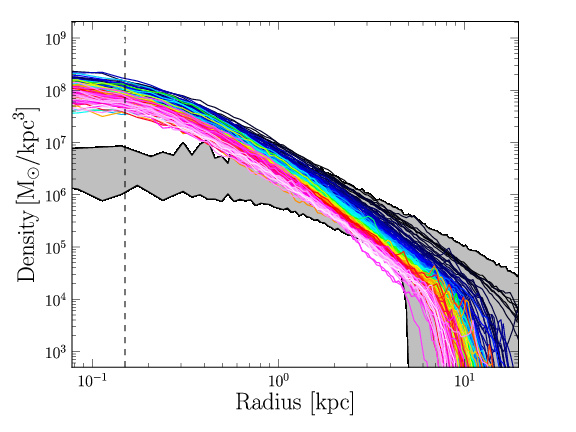}}
  \caption{The density profiles at $z=0$ of 142 dark matter haloes with masses between $10^8$ and $10^9\,\rm{M_\odot}$. The dashed
line indicates the limit of the resolution corresponding to the gravitational softening.
The gray surface show the domain coved by the pseudo-isothermal profiles (Eq.~\ref{piso}) adopted in our simulations, for the
minimum and maximum values of central densities and radii ($r_{\rm{max}}$). Only two haloes among the $144$ extracted strongly deviate from the NFW profile and 
are not seen in this plot. The reason is that at $z=0$, they both experiment a strong merging.}
  \label{fig2_condinit}
\end{figure}

(ii) Varying masses by a factor 10, from $10^8$ to $10^9\,\rm{M_\odot}$, scales the density profiles by  a factor 3-4 only
(see Fig.\ref{fig2_condinit}).


The consequences for the evolution of dSphs, are 
(i) the expansion of the universe leads to the formation of stable systems after $z=6$. 
In turn, this justifies models of dSphs in a static Euclidean space, where the expansion of the universe is neglected. 
The physics of baryons that depends on the density in physical coordinates (for example the cooling of the gas) is correct.
It would not be the case if the density in the haloes would increase with the mean density of the universe at
higher redshift.
(ii) $50\%$ of the haloes experience only minor mergers since $z=6$. Their profiles are stable and
can be modeled as isolated systems.
(iii) Although the densities of haloes with mass between $10^8$ and $10^9\,\rm{M_\odot}$ are very similar,
still they exhibit a small dispersion, a factor 3 to 4, which can help  understanding  the variety in the observed
properties of the dSph galaxies.

\subsection{Isolated systems}

The previous conclusions warrant the simulation of dSphs as isolated
systems, hence neglecting the expansion of the Universe. 
This approach is convenient in terms of CPU time, allowing to run a large number of
simulations to explore a wide range of parameters, as well as to
reach very high resolutions.

Our isolated systems are initially spherical and contain dark matter
as well as 15\% of baryons in form of gas. In a first
  approximation, we assume the same profile for the gas and the dark
  matter.

Instead of using an NFW
profile, we preferred a cored one, which is supported by observations of
normal low brightness and dwarf galaxies
\citep{blaisouellette01,deblok02,swaters03,gentile04,gentile05,
  spekkens05,deblock05,deblok08,spano08,walker11}, including the recent and high
resolution observations of the THINGS survey \citep{oh11}.  Both gas
and dark matter follow a pseudo-isothermal profile:
	\begin{equation}
	\rho_i(r)=\frac{\rho_{\rm{c},i}}{1+\left( \frac{r}{r_{\rm{c}}} \right)^2},
	\label{piso}
	\end{equation}
where $r$ is the radius, $r_c$ is the scale length of the mass
distribution, and $\rho_{\rm{c},i}$ the central mass density of the component $i$.

All parameters are set to provide profiles that are compatible with
the $\Lambda$CDM cosmological simulation discussed above, except in
the central regions, where the profiles are flattened, see the gray
area of Fig.~\ref{fig2_condinit}.

Both the initial dark matter and gas profiles are truncated at
$r_{\rm{max}}$ and the total initial mass is then defined as~:

	\begin{equation}
	M_{\rm{tot}} = 4\pi\rho_{\rm{c},\rm{tot}}\, r_{\rm{c}}^3 \left[ \frac{r_{\rm{max}}}{r_{\rm{c}}} - \arctan\left( \frac{r_{\rm{max}}}{r_{\rm{c}}} \right) \right],
        \label{Mtot}
	\end{equation}
with $\rho_{\rm{c},\rm{tot}}=\rho_{\rm{c},\rm{halo}}+\rho_{\rm{c},\rm{gas}}$.

As the precise value of $r_c$ has only a limited influence on the
evolution of the dSphs \citep{revaz09}, we fixed it to $1\,\rm{kpc}$.
$\rho_{\rm{c},\rm{halo}}$ and $\rho_{\rm{c},\rm{gas}}$ are related by
the baryonic fraction.

In the case of spherical systems and for an isotropic velocity
dispersion, the velocities of a component $i$ can be derived using the
the second moment of the Jeans equation \citep{binney87,hernquist93}~:

	\begin{equation}
        \sigma_{i}^2(r) = \frac{1}{\rho_{i}(r)}\int_r^\infty\! dr' \,\rho_{i}(r')\, \partial_{r'} \Phi(r').
	\label{siga_i}
	\end{equation}
For the halo, the velocity dispersion is directly taken from Eq.~\ref{siga_i}. For the gas, the initial 
velocities are set to zero but its initial temperature is obtained by converting the kinetic energy of Eq.~\ref{siga_i}
into thermal energy~:
	\begin{equation}
        T(r) = f_{\rm{vir}} (\gamma-1)\frac{\mu m_{\rm{H}}}{k_{\rm{B}}}\frac{3}{2}\sigma_{\rm{gas}}^2(r),
	\label{T_gas}
	\end{equation}

The parameter $f_{\rm{vir}}$ set to $0.5$ allows the gas to
        slowly flow towards the center of the system; , $m_{\rm{H}}$
        is the hydrogen mass, $\mu$, the mean atomic mass of the gas
        and $k_{\rm{B}}$ the Boltzman constant.


\section{Robustness}\label{robustness}


\subsection{Energy conservation}\label{energy_conservation}

\begin{figure}
  \resizebox{\hsize}{!}{\includegraphics[angle=0]{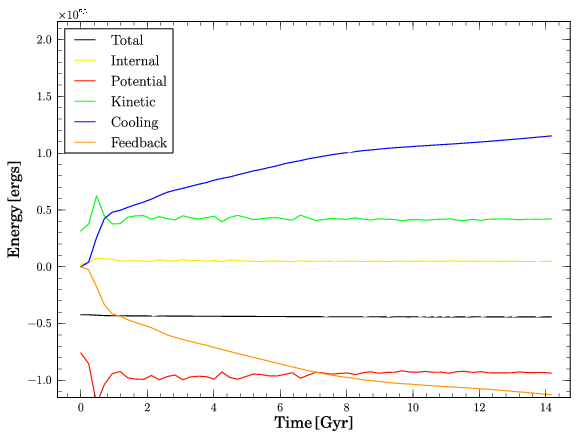}}
  \caption{Global energy budget for a dense and massive model of 
  $9.5\times 10^8\rm{M_\odot}$ ($\rho_{\rm{c,gas}}$=0.071$\rm{m_{H}/cm^3}$, $c_\star=0.05$ and 
  $\epsilon_{\rm{SN}}=0.03$) as a function of time. The total energy (in black) is the sum 
  of all other lines, namely:the total potential energy (in red), the total kinetic energy (in green), 
  the gas plus stellar internal energy (in yellow), the radiative cooling energy (in blue), and 
  the SNe feedback energy (in orange). 
  The latter is negative because it corresponds to energy injected into the system.}
  \label{fig1_energy_conservation}
\end{figure}

The first and fundamental requirement of numerical simulations is
to carefully trace all sources (gain and loss) of energy in the systems : the total potential energy, the
total kinetic energy, the gas internal energy, the radiative cooling
energy and the SNe feedback energy. 
We also include in the gas internal energy the internal energy of gas particles
converted into stars.
The precision of our energy budget
is always  better than $5\%$.  As an example,
Fig.~\ref{fig1_energy_conservation} displays the global budget of
energy of a massive $9.5\times 10^8\rm{M_\odot}$ system, with
$c_\star=0.05$ and $\epsilon_{\rm{SN}}=0.03$,
$\rho_{\rm{c,gas}}$=0.071$\rm{m_{H}/cm^3}$.
The total number of particles is $524'288$ and the softening length $25\,\rm{pc}$.
Such a high mass system,
with a high central mass density, experiences star formation sustained
at a high level (see Section~\ref{regimes_of_starformation}), making
it a difficult case from a numerical point of view.  Nevertheless, the
energy is conserved within $5\%$. This is remarkable with regards to
the long integration time ($14\,\rm{Gyr}$) compared to the dynamical
time of the system (about $50\,\rm{Myr}$).
Fig.~\ref{fig1_energy_conservation} also reveals a clear balance
between the injected energy by the SNe explosions and the radiative
cooling.  This means that although the cooling is switched off during
an adiabatic period $t_{\rm{ad}}$ (see Section~\ref{feedback}), a
large fraction of the feedback energy is still radiated away.

\subsection{Convergence}\label{convergence_tests}

We checked the convergence  of our results with increasing resolution,
i.e.,  the number of  particles   was increased and  the  gravitational
softening $\epsilon_{\rm{g}}$  was decreased accordingly. For  this, we
performed  height  sets of simulations  on  systems  with two different
masses, $3$ and $9.5\times  10^8\rm{M_\odot}$, xploring a weak 
($\epsilon_{\rm{SN}}=0.03$) and
a strong ($\epsilon_{\rm{SN}}=1$ feedback). The sizes of the systems
are the same and their central gas density are respectively
$\rho_{\rm{c,gas}}=0.025$ and
$\rho_{\rm{c,gas}}=0.066\,\rm{m_{H}/cm^3}$.

\begin{figure}
  \resizebox{\hsize}{!}{\includegraphics[angle=0]{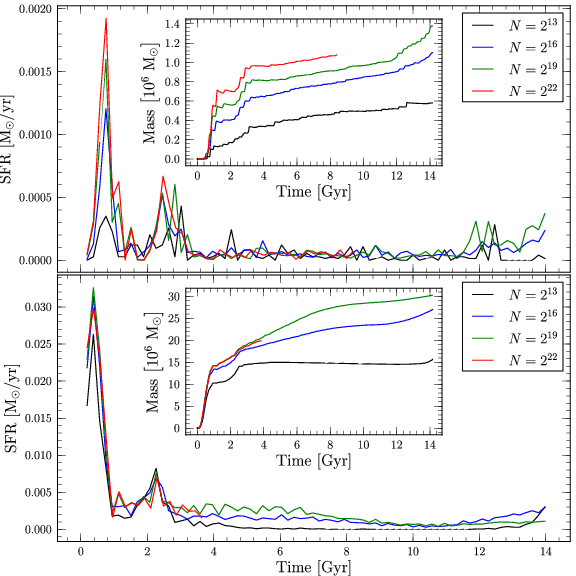}}
  \caption{Star formation rate and stellar mass as a function of time for the two models of Table
\ref{table_convergence_tests}.   Each line corresponds to a different resolution.}
  \label{fig1_convergence_tests}
\end{figure}

The initial number of particles went from $2^{13}=8'092$ to
$2^{22}=4'194'304$, corresponding to a mass resolution of about
$10^4\,\rm{M_\odot}$ and $20\,\rm{M_\odot}$, respectively. This is
equivalent to changing the mass resolution by a factor 512, and the
spatial resolution by a factor 8. We start by discussing the case of the weak feedback.
The main properties of the
simulations are summarized in Tab.~\ref{table_convergence_tests}; for all, $c_\star=0.05$ and
$f_{\rm{vir}}=0.25$.   
Fig.~\ref{fig1_convergence_tests} displays the
evolution of the star formation rates and the corresponding increases
in stellar mass.  The amount of stars formed results from the complex
interplay between the gas cooling and feedback heating and thus traces
well all the implemented physical
processes. Fig.~\ref{fig1_convergence_tests} shows that changing the
resolution keep the peaks of star formation at the same epochs.
High resolutions models form two to five times more stars
over a longer period than the low resolution ones.  This is the
consequence of an improved sampling of the high central density peaks,
which translates into shorter cooling times.

\begin{table}
\begin{minipage}[t]{\columnwidth}

\caption[]{Properties of the runs performed for the convergence tests.
$M_{\rm{tot}}$ is the initial total mass of the system,
$\rho_{\rm{c,gas}}$ is the initial central gas density,
$N$ is the initial total number of particles (gas and dark matter) and
$\epsilon_{\rm{g}}$ is the gravitational softening. These models were
run with $c_\star=0.05$ and $\epsilon_{\rm{SN}}=0.03$

}
\label{table_convergence_tests}

\centering
\renewcommand{\footnoterule}{}

  \begin{tabular}{c c l r r}
    \hline\hline
      $M_{\rm{tot}}$
& $\rho_{\rm{c,gas}}$
& N
& $m_{\rm{gas}}$
& $\epsilon_{\rm{g}}$ \\
      $[10^8\,\rm{M_\odot}]$ & [$\rm{m_{H}/cm^3}$] & & $[\rm{M_\odot}]$ & $[\rm{pc}]$ \\
    \hline\hline

9.5 &0.071 & $2^{13}=8'192$ &    $3.16\times 10^4$ & 100\\
- &- & $2^{16}=65'536$ &    $4.11\times 10^3$ & 50\\
- &- & $2^{19}=524'288$ &    $5.00\times 10^2$ & 25\\
- &- & $2^{22}=4'194'304$ &    $6.17\times 10^1$ & 12.5\\
   \hline
3.0 &0.022 & $2^{13}=8'192$ &    $1\times 10^4$ & 100\\
- &- & $2^{16}=65'536$ &       $1.3\times 10^3$ & 50\\
- &- & $2^{19}=524'288$ &    $1.58\times 10^2$ & 25\\
- &- & $2^{22}=4'194'304$ &    $1.95\times 10^1$ & 12.5\\
    \hline
  \end{tabular}

\end{minipage}
\end{table}

The impact of a change in numerical resolution is slightly larger for
small systems than for massive ones.  The two simulations with
$4'194'304$ particles are still running at the time of writing this
paper hence their corresponding red curves are still incomplete.
However, during the first $4$  $\rm{Gyr}$, the star formation history
of the $9.5\times 10^8\rm{M_\odot}$ model is similar for the
$N=2^{19}$ and $N=2^{22}$ resolutions. 
In the case of the smaller mass
system, the difference in stellar mass between the $N=2^{19}$ and
$N=2^{16}$ models is of the order of $15$\% during the $8$ first
$\rm{Gyr}$. This increase in mass is nevertheless moderate and does not translate
into different star formation histories, meaning different  age or metallicity
distributions and chemical abundance patterns.

\begin{figure}
  \resizebox{\hsize}{!}{\includegraphics[angle=0]{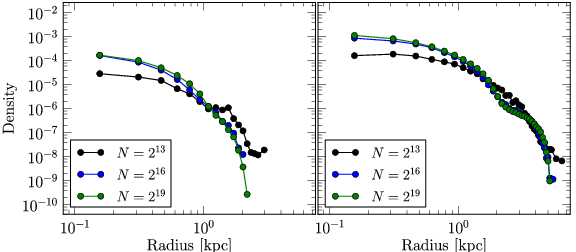}}
  \caption{Stellar density profiles for the models of mass $3$ (left) and $9.5\times 10^8\rm{M_\odot}$ (right)
  described  in Table \ref{table_convergence_tests} with different resolutions, at $t=14\,\rm{Gyr}$.}
  \label{fig3_convergence_tests}
\end{figure}

Fig.~\ref{fig3_convergence_tests} show the stellar profiles, which depend directly on the dynamics.
They show that a low spatial resolution leads to flattening of the profiles compared  
to higher resolution.

\begin{figure}
  \resizebox{\hsize}{!}{\includegraphics[angle=0]{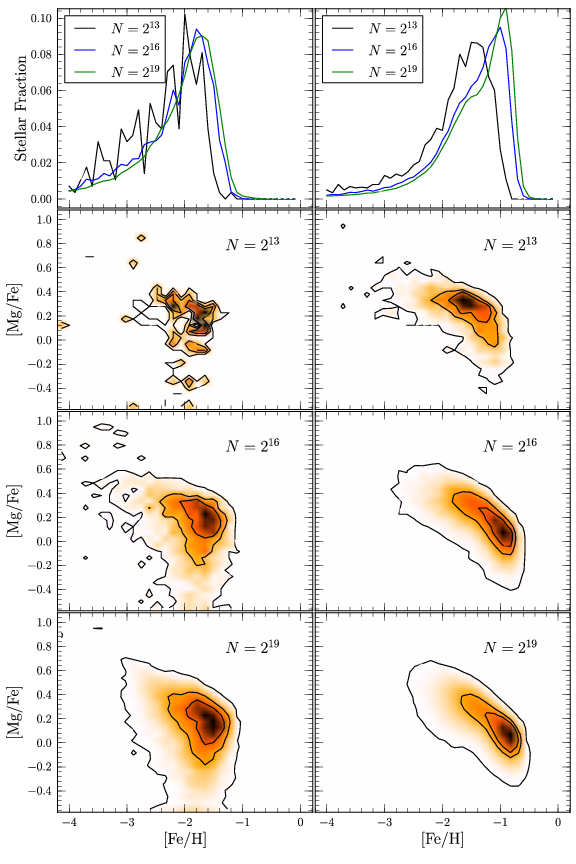}}

   \caption{[Fe/H] distributions and [Mg/Fe] vs [Fe/H] diagrams at
     $t=14\,\rm{Gyr}$ for the two models described in Table
     \ref{table_convergence_tests}. The galaxy with a mass of $3\times
     10^8\rm{M_\odot}$ is shown on the left column, while the model
     with $9.5\times 10^8\rm{M_\odot}$ is displayed on the right
     column.}
  \label{fig2_convergence_tests}
\end{figure}

Indeed, Fig.~\ref{fig2_convergence_tests} illustrates how  the metallicity
distributions and abundance ratios are sensitive to the resolution.
A low a resolution ($N<2^{13}$) tends to shift artificially the
metallicity distribution to lower values and only deliver an
incomplete view of the chemical evolution.  Satisfying convergence is
again obtained for $N=2^{16}$ particles,  leading to [Fe/H]
differences of less than $0.1$ dex.

  We performed the same tests for a strong feedback ($\epsilon_{\rm{SN}}=1$)
  and could verify that convergence is also satisfied, despite the
  fact that for these extreme feedback cases, inhibiting star
  formation, the resulting low number of stars formed may lead to
  large noise as star formation is implemented as a stochastic
  process.

In summary, our tests demonstrate the reliability of the numerical
implementation of the physical processes described in
Section~\ref{code_gear}, which are independent of the
number of particles used. Above a resolution of $N=2^{16}$  particles, the physical
quantities are described with sufficient accuracy. The convergence arises from the balance
between feedback and cooling; our systems are self-regulated.
In the rest of this paper, we use $N=2^{16}$, a good trade-off between
resolution and CPU time.

\subsection{Random number seed}\label{random_number_seed}

Whilst convergence of our results is obtained, nevertheless our
systems could still be nonlinear at fixed resolution, due to the
complexity of the physics.  For example, changing the cooling impacts
the star formation which in turn impacts the feedback and the cooling
again, etc. The integration time of $14\,\rm{Gyr}$ is long with
regards to the dynamical times of the galaxies, of the order of $100\,\rm{Myr}$. 
Small perturbations might produce a large dispersion in the final properties of the
galaxies.  In the following, we investigate this issue and evaluate
the error bars attached to our outputs.  Random numbers are used by
the star formation algorithm and can act as perturbations.  Therefore,
we performed three different sets of simulations, in which only the
random number seed was modified.  The left part of
Table~\ref{table1_random_number_seed} summarizes the parameters of our
simulations.

\begin{table*}

\caption[]{Properties of the models for which the random number seed was modified.}
\label{table1_random_number_seed}

\centering
\renewcommand{\footnoterule}{}

  \begin{tabular}{l c c c c | c c c}
    \hline\hline
    model 
    & $M_{\rm{tot}}$  
    & $\rho_{\rm{c,gas}}$ 
    & $c_\star$ 
    & $\epsilon_{\rm{SN}}$ 
    & $\delta(M_\odot)$
    & $\delta(L_{\rm{v}})$
    & $\delta(\rm{[Fe/H]})$\\
    & [$10^8\,\rm{M_\odot}$] 
    & [$\rm{m_{H}/cm^3}$] 
    & 
    & 
    & [\%]
    & [\%]
    & [\%]\\
    \hline\hline                  
    885		&3	& 0.022	&	0.05		& 0.03	& 4 &  9 & 11\\
    1111	&8	& 0.029	&	0.1		& 0.05	& 6 & 26 & 6\\
    1056	&7	& 0.059	&	0.03		& 0.03	& 3 & 13 & 0\\
    \hline           
  \end{tabular}   

\end{table*}

Fig.~\ref{fig0_appendix_2} displays the case of model \#1111, which
presents the largest deviations.  They  are anyway very
small. Even after several dynamical times of evolution,
the system is not strongly affected by the perturbations generated
by the use of a different random number sequence.

Table~\ref{table1_random_number_seed} quantifies the relative
variations in stellar mass, V-band luminosity, and final mode
metallicity. These variations are estimated as the ratio between the
absolute difference between the minimum and the maximum of the
measured quantity divided by its mean value.  The final stellar masses
deviate by $\sim 6\%$ at most and the metallicities by $11\%$, while
the luminosities can vary by  $26\%$, due to the strong
dependence of the luminosity on the stellar age. These
values can be interpreted as the error bars intrinsic to the models. They 
must be kept in mind when drawing comparison between models and
observations.


\section{Parametrization}\label{parameter_effect}


Physical processes involved in galaxy formation, such as star
formation, feedback or cooling occur typically on lengths and time
scales much smaller than the ones resolved by current numerical
simulations.  As described in Section~\ref{code_gear}, such sub-grid
physics is treated through phenomenological prescriptions and requires
the introduction of a set of parameters.  Table~\ref{parameters}
compiles the list of our model parameters.  In the following, we
explore their role. Unless specified otherwise, the discussions are based
on the full set of 393 simulations.

\begin{table*}
  \caption{
  List of model parameters. The reference values indicate our final set of adopted parameters 
  based on the four generic models of Local Group dSphs. The last column indicates whether or not 
  the parameters were varied (see Section~\ref{parameter_effect} and Section~\ref{mass_profile}), 
  in which case we give their range of values.
  }
  \begin{center}
  \begin{tabular}{l l c c c r}
     \hline\hline       
     context                    &       quantity                        &       symbol                  &       reference values         &       variation\\ 
     \hline\hline
      initial conditions        &       baryonic fraction               & $f_{\rm{b}}$                  &       $0.15$                  &       no      \\
                                &       dark halo core radius           & $r_c$                         &       $1\,\rm{kpc}$           &       no      \\
                                &       virial fraction                 & $f_{\rm{vir}}$                &       $0.5$                   &       no      \\
                                &       total mass                      & $M_{\rm{tot}}$                &       -                       &       [$1-9 \times 10^8\,\rm{M_\odot}$]     \\
                                &       gas central density                 & $\rho_{\rm{c},\rm{tot}}$      &       -                       &       [0.005 - 0.063$\,\rm{m_{H}/cm^3}$ ]    \\
                                &       initial outer radius            & $r_{\rm{max}}$                &       -                       &       [1.91 - 24.4 kpc]     \\
      \\
      gravity                   &       softening                       & $\epsilon_{\rm{g}}$           &       $50\,\rm{pc}$           &       [adaptive]     \\
      \\
      sph                       &       number of neighbors             & $N_{\rm{ngb}}$                &       50                      &      [50, 100]     \\
      \\
      cooling                   &       cooling function                & $\Lambda$                     &       \citet{sutherland93,maio07}                &       no      \\
      \\
      star formation            &       parameter                       &  $c_\star$                    &       $0.05$                  &       [0.01 - 1]    \\
                                &       critical density               &  $\rho_{\rm{sfr}}$            &       $0.1\,\rm{m_{H}/cm^3}$  &       [0.05 - 100$\,\rm{m_{H}/cm^3}$]     \\
                                &       maximal number of stellar       \\
                                &       particles per gas particle      &  $N_\star$                    &       $4$                     &       [1 - 15]     \\
                                &       gas temperature                 &  $T_\star$                    &       $3 \times 10^4\,\rm{K}$  &       no      \\
\\
      stellar evolution         &       initial mass function           &   IMF                         &       \citet{kroupa01}        &       [Kroupa,Salpeter]     \\
                                &       yields                          &   -                           &       \citet{iwamoto99,tsujimoto95}&  no      \\

\\
      thermal feedback          &       supernova energy                &   $E_{\rm{SN}}$               &       $10^{51}\,\rm{ergs}$    &       no     \\
                                &       efficiency                      &   $\epsilon_{\rm{SN}}$        &       $0.03$                  &       [0.01 -1]     \\
                                &       adiabatic period                &   $t_{\rm{ad}}$               &       $5\,\rm{Myr}$           &       [2 - 30 Myr]     \\

    \hline
    \hline
  \end{tabular}
  \label{parameters}
  \end{center}
\end{table*}

\subsection{Feedback efficiency $\epsilon_{\rm{SN}}$}\label{esn}

\begin{figure}
  \resizebox{\hsize}{!}{\includegraphics[angle=0]{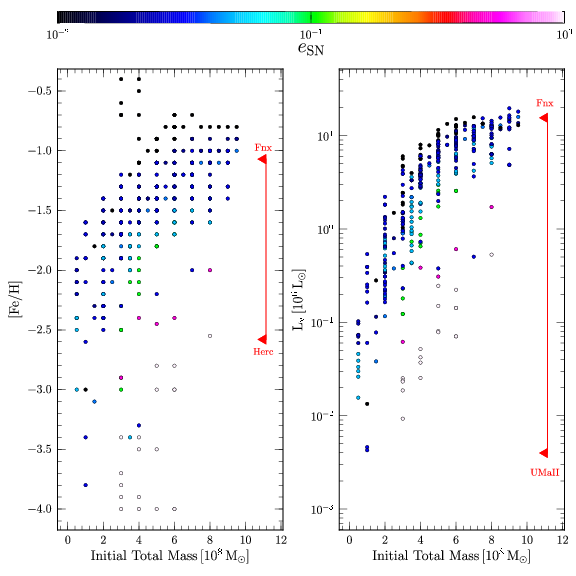}}
  \caption{Final \mfe\ and luminosity $L_{\rm{v}}$ as a function of the total
    initial mass of the system. Our full set of simulations is shown. The colors code the
    supernova feedback efficiency, between 0.01 and 1. In each panel,
    a red line indicate the range of metallicity or luminosity covered
    by the Local Group dSphs.  }
  \label{fig1_supernova_efficiency}
\end{figure}

The feedback efficiency, $\epsilon_{\rm{SN}}$, is the fraction of
energy released by the explosion of supernovae, which is effectively
deposited into the interstellar medium. {\rm The remaining fraction is
  assumed to be radiated away, without impacting the system.}  Our
feedback prescription is thermal (see Section~\ref{feedback}), hence
it modifies the temperature, the pressure and consequently the density
of the gas surrounding the exploding stellar particles and in turn
impacts the eligibility of the gas particles to further form stars.
Fig.~\ref{fig1_supernova_efficiency} displays the final mode stellar
metallicity \mfe and V-band luminosity ($L_v$) as a function of the
initial total mass of the system.

If 100\% of the supernova energy is injected in
the ISM ($\epsilon_{\rm{SN}}=1$, open circles in
Fig~\ref{fig1_supernova_efficiency}), the final mode metallicity of
the system is always below the values measured for the Local Group dSphs and
UFDs (red vertical line).  Similarly the luminosities are generally
very low.  For such high values of $\epsilon_{\rm{SN}}$, the feedback
energy is maximal and strongly heats and dilutes the ISM pushing the
gas particles away from the criteria for star formation, which is
stopped.  The subsequent increase of the cooling time, mainly driven
by the decrease of the gas density, is so strong that it inhibits any
star formation during the rest of the galaxy evolution.

Clearly $\epsilon_{\rm{SN}}$ must be decreased below 1 in order to
reach metallicities and luminosities compatible with the observations.
Fig.~\ref{fig1_supernova_efficiency} qualifies
$\epsilon_{\rm{SN}}$ in the range $0.1$ to $0.01$.   Besides, it also shows that the
initial total mass of the systems drives the final luminosity and
metallicities. $\epsilon_{\rm{SN}}$ primarily influences the least
massive systems.  At very low $\epsilon_{\rm{SN}}$ the cooling of the
gas is no longer counterbalanced by the supernovae
feedback: stars can form efficiently  and longer, leading to high 
metallicities.   Fig.~\ref{fig1_appendix_2} and
\ref{fig2_appendix_2} illustrate also how the ISM becomes more
homogeneous with decreasing $\epsilon_{\rm{SN}}$,  diminishing the dispersion in stellar abundance
ratios. To further constrain $\epsilon_{\rm{SN}}$, keeping its
value fixed for all galaxies, requires to consider the full set of dSph
properties.  Our generic models (see Section~\ref{generic_models})
are obtained with $\epsilon_{\rm{SN}}=0.03$.

%

\subsection{Threshold density for star formation $\rho_{\rm{sfr}}$}\label{rho_sfr}

The threshold density for star formation, $\rho_{\rm{sfr}}$, has been
introduced in a cosmological context by \citet{summers84} in order to
avoid the formation of stars in low density regions.  The value of
$0.1\,\rm{m_{H}/cm^3}$, that is widely used in the literature
\citep[e.g.,][]{alimi03,stinson06,valcke08}, was introduced by \citet{katz96}
and corresponds to the mean density of the warm neutral medium of our
galaxy.  Since $\rho_{\rm{sfr}}$ fixes the local density at which
stars are allowed to form, it may as well influence the global galaxy
star formation history.

The importance of this threshold density has been discussed in
the context of disk galaxies \citep{tasker06,tasker08}. \citet{saitoh08}
concluded that only models using a high threshold density ($\rho_{\rm{sfr}} =100\,\rm{m_{H}/cm^3}$) are
able to reproduce the complex, inhomogeneous and multiphase structure of the ISM.
Hereafter, we investigate the consequences of varying the critical
density in the context of dSph galaxies.

\begin{figure}
\resizebox{\hsize}{!}{\includegraphics[angle=0]{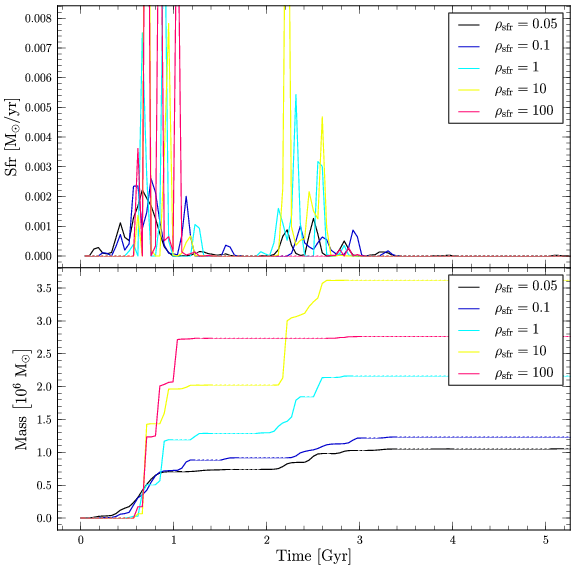}}
\caption{The effect of varying $\rho_{\rm{sfr}}$. This figure illustrates the case of
a $3.5\times 10^8\rm{M_\odot}$ mass system, with an initial gas density of $0.025\,\rm{m_{H}/cm^3}$.
The upper panel shows the evolution of the \sfr with time, while the lower panel displays how mass increases. 
}
\label{fig2_rhostar}
\end{figure}

For $c_{\star}=0.05$ and $\epsilon_{\rm{SN}}=0.05$, we varied $\rho_{\rm{sfr}}$ from
$0.05$ to $100\,\rm{m_{H}/cm^3}$ for two different initial masses,
$3.5$ and $7\times 10^8\,\rm{M_\odot}$, within
$r_{\rm{max}}=8\,\rm{kpc}$ and the initial central gas
densities of $0.025$ and $0.053\,\rm{m_{H}/cm^3}$, respectively.  A zoom on the first
few Gyr evolution of the star formation rate and stellar mass of the
$3.5 \times 10^8\,\rm{M_\odot}$ model is provided in
Fig.~\ref{fig2_rhostar}. It shows that the increase of $\rho_{\rm{sfr}}$
imposes an increasing delay for the onset of the star formation, because the
gas needs to be gradually denser. Moreover, one passes from a regime
of large and moderately intense peaks of star formation to a series of higher intensity
and higher frequency episodes, creating a 
larger  number of stars, although varying $\rho_{\rm{sfr}}$
by a factor $1000$ does not change the final stellar mass by more than
a factor $2.5$. Drastic changes in the metallicity distributions
and abundance ratios can be witnessed in Fig.~\ref{fig7_appendix_2}.

Following the course of the formation of the galaxies, it stands out
that for low $\rho_{\rm{sfr}}$ ($<1\,\rm{m_{H}/cm^3}$), the first
generations of stars form within a relatively large region (radius
$\sim 300\,\rm{pc}$), roughly corresponding to the SPH sphere (radius
$\sim 250\,\rm{pc}$) in these low density medium.  Metals and energy
feedback are homogeneously distributed.  For larger star formation
density thresholds, the evolution of the systems is quite
different. Because the density of the gas is higher, the size of the
star forming region is smaller, with typical radii of the order of
$\sim 30\,\rm{pc}$ for $100\,\rm{m_{H}/cm^3}$, again corresponding
roughly to the SPH sphere radius ($\sim 15\,\rm{pc}$).  The mechanism
of metals and energy feedback being linked to the SPH formalism,
metals are distributed very locally.  The release of the energy of the
supernovae explosions creates bubbles. Those inflate and push back the
surrounding gas, which significantly increases its density.  The rims
of the gas bubbles are dense enough to cool quickly.  They become
unstable and form stars. In this way, star formation induces star
formation.  This explains the increase of stellar mass for large
$\rho_{\rm{sfr}}$, observed in the bottom of Fig.~\ref{fig2_rhostar}.

The metallicity distributions and abundances ratios generated by
$\rho_{\rm{sfr}}$ $>1\,\rm{m_{H}/cm^3}$ look incompatible with the
observed galaxy properties and are consequently discarded.  We thus
keep the value of $0.1$ which best fit the abundances constraints.

\subsection{Star formation parameter \cstar}\label{cstar}

The star formation parameter \cstar is a dimension-less parameter that
controls the amount of stars formed from the gas.  In the context of
spiral galaxies, \cstar may be calibrated using a Kennicutt-Schmidt
law \citep{schmidt59,kennicutt98}. Unfortunately, as dSphs are devoid
of gas, no similar relation exists for them.  The impact of the
variation of \cstar has been discussed by \citet{stinson06} in the
context of their feedback blast wave model.  Their mean star formation
rate varies by only a factor $\sim 1.2$ when \cstar goes from
$0.05$ to $1$.

\begin{figure}
   \resizebox{\hsize}{!}{\includegraphics[angle=0]{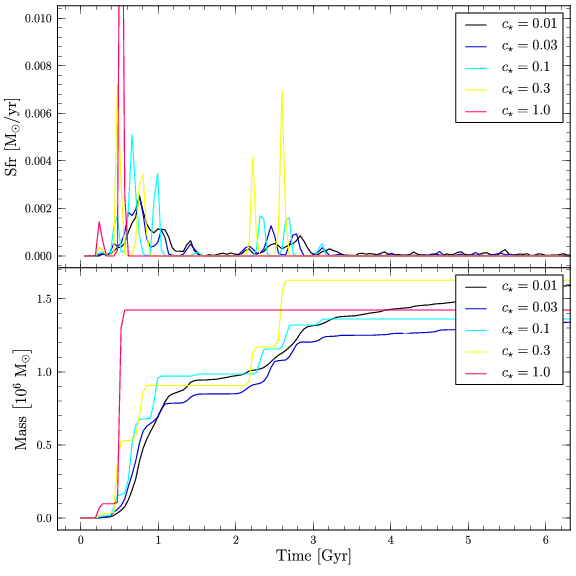}}
   \caption{Star formation rate and stellar mass as a function of time for models of mass $3.5\times 10^8\rm{M_\odot}$ ($\rho_{\rm{c,gas}}=0.025\,\rm{m_{H}/cm^3}$) with different \cstar.}
   \label{fig_cstar_1}
\end{figure}

We performed a similar study for our pressure supported systems,
varying \cstar by two dex, from $0.01$ to $1$. We considered two
models with $3.5\times10^8\rm{M_\odot}$
($\rho_{\rm{c,gas}}=0.025\,\rm{m_{H}/cm^3}$, $\epsilon_{\rm{SN}}=0.05$) and
$7\times10^8\rm{M_\odot}$
($\rho_{\rm{c,gas}}=0.053\,\rm{m_{H}/cm^3}$, $\epsilon_{\rm{SN}}=0.05$).  
The comparison of the galaxy
final state is provided in Fig.~\ref{fig5_appendix_2} and
\ref{fig6_appendix_2}.  Fig.~\ref{fig_cstar_1} offers a zoom into the
first few Gyr of evolution for the $3.5\times10^8\rm{M_\odot}$ model.
Independent of the mass of the dSph, large \cstar results in a very
discontinuous star formation history. This is a direct consequence of
the coupling between star formation and feedback.  In the extreme case of
$c_\star=1$ and $M_{\rm{tot}}=3.5\times10^8\rm{M_\odot}$, 99\% of the
stars are formed during a first burst. This dramatically impacts the
metallicity distribution and metal abundances (see
Fig.~\ref{fig5_appendix_2}). Metal-poor stars ($\rm{[Fe/H]}<-2.5$)
with high $\rm{[Mg/Fe]} \sim 1$ ratio are dominant and make these
systems unrealistic. The long period of quiescence following the first
and dominant peak results from an important physical process, which
will be discussed in Section~\ref{fates_for_feedbacks}.

On the contrary, at lower \cstar the galaxy experiences a more
continuous star formation history, resulting from a better
self-regulation between cooling and heating.  The moderate
intensity of the initial bursts allows the gas to cool. This gives rise
to the presence of intermediate age stellar populations and higher
mean metallicity.

The full evolution of the stellar mass with time is displayed in
Fig.~\ref{fig5_appendix_2} and \ref{fig6_appendix_2}.
Differences in the final
stellar masses remain smaller than $50\%$. Increasing \cstar slightly
decreases the quantity of stars formed. This counter intuitive result
is also found by \citet{stinson06} at large \cstar.
Indeed, a model with a higher \cstar forms quickly a large quantity of stars. 
The  feedback energy due to the SNeII explosions is released nearly instantaneously. The
gas is blown out and the star formation is quenched.
At later times, the large number of  SNeIa continue to prevent star formation, as
discussed in more details in Section~\ref{fates_for_feedbacks}.

\subsection{Number of stellar particles $N_\star$ created from a gas particle}

The mass of the stellar particles $m_{\star}$ is linked to the mass of the gas
particles $m_{\rm{g}}$ via the parameter $N_\star=m_{\rm{g}}/m_{\star}$.
It can also be seen as the maximum number of stellar particles that
can be created per gas particle in absence of accretion of gas.
In principle, Eq.~\ref{pstar} regulates the impact of the resolution
in mass of the simulations, leading to the same integrated star formation
rate. Nevertheless, this is not fully the case once
cooling and feedback are taken into account.
Augmenting $N_\star$ increases the sampling of the star
formation.  A small number of big particles are replaced by a larger
number of smaller ones, distributed along the same period of time, and for
an equal final stellar mass.  The price to pay is obviously a large
increase of the CPU time needed for a simulation.  According to
\cite{springel03}, a good compromise between time resolution and CPU
saving is $N_\star=4$.

We varied $N_\star$ between $1$ and $15$ for a $5\times10^8\rm{M_\odot}$
($r_{\rm{max}}=8\,\rm{kpc}$, $\rho_{\rm{c,gas}}=0.037\,\rm{m_{H}/cm^3}$, $c_{\star}=0.025$, $\epsilon_{\rm{SN}}=0.02$). 
The results are displayed in
Fig.~\ref{fig9_appendix_2}.  Two effects are observed. The final mode
stellar metallicity of the systems decreases within larger $N_\star$, by a factor
two passing from $N_\star$=1 to 15.  A large number of small stellar
particles regularly injects feedback energy in the interstellar
medium, preventing it to cool and be eligible to star formation again.
Conversely, as the probability of forming stars for large $N_\star$ is
small, the system is slightly cooler and denser when the first star form,
leading to a higher star formation peak. Whilst the mean metallicity
of the systems are similar within $0.3$ dex, a clear trend is observed in
the abundance ratios. The dispersion in [Mg/Fe] 
is significantly reduced for higher $N_\star$ (high chemical mixing),  again owing to the
frequent feedback of the numerous stellar particles, starting early in
the evolution of the galaxies.

Small abundances dispersions at low metallicities are the rule in observed galaxies.
This favors high $N_\star$. The dispersion does not increase substantially
between $N_\star=4$ and  $N_\star=15$ (see Fig.~\ref{fig9_appendix_2}),
therefore saving CPU time, we set it to 4.

\subsection{Supernova adiabatic period $t_{\rm{ad}}$}\label{tad}

The adiabatic period $t_{\rm{ad}}$ fixes the period of time during
which the cooling of the gas particles, that have received the thermal
supernova feedback, is stopped.  As discussed in
Section~\ref{feedback}, this period improves the feedback efficiency
by avoiding the instantaneous loss of energy by radiative cooling.
$t_{\rm{ad}}$ must be much larger than the dynamical time steps, which
is at most $0.2\,\rm{Myr}$ in our case as imposed by the resolution of
the chemical evolution (see Section~\ref{stellar_ejecta}), in order to
ensure the convergence of the simulation. It is often set to
$30\,\rm{Myr}$, the shortest lifetime of the least massive SNII
progenitor ($8\,\rm{M_\odot}$).  \citet{stinson06} proposed a new
feedback recipe based on an analytical treatment of supernova
blast-waves.  In this model, $t_{\rm{ad}}$ depends on the local gas
density and pressure.

We explored the effect of $t_{\rm{ad}}$, between $2$ and $30\,\rm{Myr}$,
for two models, $M_{\rm{tot}}=3.5$ ($r_{\rm{max}}=8\,\rm{kpc}$, $\rho_{\rm{c,gas}}=0.025\,\rm{m_{H}/cm^3}$) 
and $7\times 10^8\,\rm{M_\odot}$ ($r_{\rm{max}}=8\,\rm{kpc}$, $\rho_{\rm{c,gas}}=0.051\,\rm{m_{H}/cm^3}$). 
Both models have $c_{\star}=0.05$ and $\epsilon_{\rm{SN}}=0.05$.
The results are displayed in Fig.~\ref{fig3_appendix_2} and \ref{fig4_appendix_2}.
Increasing $t_{\rm{ad}}$ decreases the final stellar mass in both
cases.  This is the consequence of a lower cooling.  However, the
effect remains weak. After $14\,\rm{Gyr}$, the amount of the stellar
mass formed is diminished by a factor of 1.3 to 1.6.
This is not sufficient to change substantially the chemical properties of the final dwarf.
In the following, we fix $t_{\rm{ad}}$ to $5\,\rm{Myr}$.

\subsection{Number of particles in a softening radius $N_{\rm{ngb}}$}\label{nngb}

As described in Section~\ref{stellar_ejecta} and \ref{feedback}, the
stellar ejecta as well as the feedback energy are spread in the SPH
sphere.  Increasing the size of the SPH sphere, i.e., increasing the
number of neighbors when computing the local physical
quantities, allow to distribute metals at larger radii.

In galaxy cluster simulations, \citet{tornatore07} varied
$N_{\rm{ngb}}$ from $16$ to $128$.  They reported only marginal
differences, although increasing the number of neighbors produced
somewhat higher star formation rates, presumably due to the largest number
of  particles affected by metal-line cooling.

We performed a similar test with  a $4\times
10^8\,\rm{M_{\odot}}$ ($r_{\rm{max}}=8\,\rm{kpc}$,
$\rho_{\rm{c,gas}}=0.029\,\rm{m_{H}/cm^3}$) simulation, and changing
$N_{\rm{ngb}}$ from $50$ to $100$.  The final state of the systems are
display in Fig.~\ref{fig12_appendix_2}. The difference between the two
models is very small, in agreement with \citet{tornatore07}.  However,
we reach an opposite conclusion. When doubling $N_{\rm{ngb}}$, the
final stellar content is slightly decreased (factor $1.33$). This
induces a slight decrease of the mean metallicity ($<0.1$ dex).  The
reason is probably linked to the different treatment of the feedback
in the two studies. \citet{tornatore07}'s model is based on the wind
scheme of \cite{springel03}.

\subsection{The IMF}

\begin{figure}
\resizebox{\hsize}{!}{\includegraphics[angle=0]{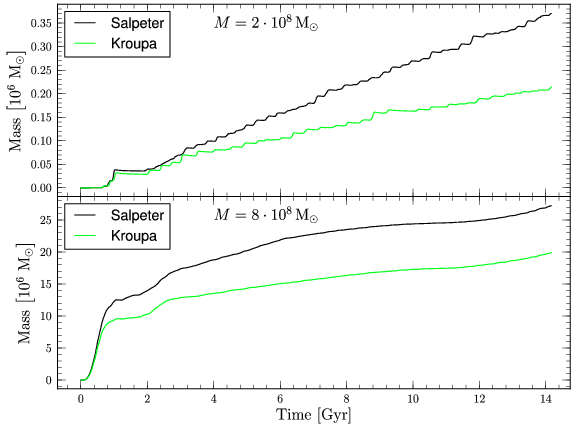}}
\caption{The difference in stellar mass growth with time when changing of IMF. Two examples are shown, at 
$2 \times 10^8 \rm{M_{\odot}}$ ($\rho_{\rm{c,gas}}=0.015\,\rm{m_{H}/cm^3}$)
and $8\ \times 10^8\rm{M_{\odot}}$  ($\rho_{\rm{c,gas}}=0.053\,\rm{m_{H}/cm^3}$).}
\label{fig_imf}
\end{figure}

We finally checked the impact of the choice of IMF on our results.
For this, we used the classical Salpeter IMF \citep{salpeter55}, defined by
a constant  slope of $-2.35$ . The \citet{kroupa01}'s IMF contains
less stars with masses smaller that $0.2\,\rm{M_{\odot}}$ than the
Salpeter's one.  As a consequence of normalization, this means more
stars in the mass range of supernovae ($>8\,\rm{M_{\odot}}$) by a
factor 1.5, and thus more feedback energy. Fig.~\ref{fig_imf}
illustrates that less stars are formed with a Kroupa IMF than with a
Salpeter one.

This weak difference in the final stellar mass does not impact substantially neither the metallicity distributions nor
the [Mg/Fe] abundance ratios (see Fig.~\ref{fig10_appendix_2} and \ref{fig11_appendix_2}).

\section{Mass and central density}\label{mass_profile}

In the previous Section, we explored the role of the parameters
inherent to \texttt{GEAR}.  We now discuss the effect of the initial
total mass of the galaxy and of the initial central gas density, which
are  linked to the initial conditions.

\citet{revaz09} found the galaxy total initial mass estimated within a
fixed radius to be a dominant parameter, increasing both the final
metallicity and the luminosity of the galaxies.  We confirm this result as
illustrated in Fig.~\ref{fig1_supernova_efficiency}, which can be
compared with the Fig.~5 of \citet{revaz09}.

As described in Section~\ref{dmh}, the dark matter profiles issued
from the $\Lambda$CDM simulations exhibit a small dispersion in their
central regions.  This deserves investigation.  Indeed, the
initial gas profile follows the dark matter one in our models, and
the gas density determines the cooling time.  Moreover, the
star formation criterion involves a threshold in gas density.  All
these facts motivated us to examine the role played by the central
density in addition to that of the total mass.  In a previous publication,
\cite{carraro01} had observed different star formation histories
for objects of the same total mass, when their the collapse phase
started at different initial densities.

\begin{figure}
  \resizebox{\hsize}{!}{\includegraphics[angle=0]{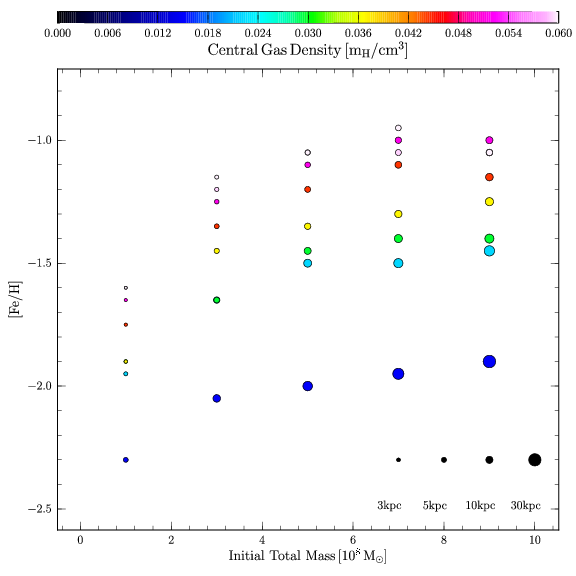}}
  \caption{Final mode \mfe\  as a function of the total initial mass of the system.
  The size of the circles scales with the initial size (radius) of the system, as featured at the bottom right
  of the figure. The color code the  initial gas density.  For all models
  $c_{\star} = 0.05$ and $\epsilon_{\rm{SN}}$=$0.03$.}
  \label{fig1_central_density}
\end{figure}

At a given central density ($\rho_{\rm{c,gas}}$), the total mass
($M_{\rm{tot}}$) of the system can be varied by increasing/decreasing
its size with $r_{\rm{max}}$.  Similarly, for a given total mass, the
central density can be varied increasing/decreasing its size with
$r_{\rm{max}}$.  Fig.~\ref{fig1_central_density} presents the
variation of the final mode metallicity as a function of the initial
mass and the central gas density.  We present $40$ simulations - all
with $c_\star=0.05$ and $\epsilon_{SN}=0.03$ -, where
$\rho_{\rm{c,gas}}$ is varied from $0.007$ up to
$0.063\,\rm{m_{H}/cm^3}$ and $M_{\rm{tot}}$ from $1$ to $9
\times 10^8\,\rm{M_{\odot}}$. Consequently, the outer galaxy radius
varies between $1.8$ to $25\,\rm{kpc}$.

For the range of metallicities covered by the Local Group dSphs, an
increase of the mass by a factor 10, at fixed central density,
increases the final metallicity of the system by only a few tenth of
dex. The largest variation is obtained for very small systems:
from $10^8\,\rm{M_{\odot}}$ to $3 \times 10^8\,\rm{M_{\odot}}$, the
mode metallicity can increase by $0.4\,\rm{dex}$ at fixed density; above $3
\times 10^8\,\rm{M_{\odot}}$, variations soften until they are
essentially indiscernible. Above $3 \times 10^8\,\rm{M_{\odot}}$, the
additional amount of matter provided by the extension of the system
(increase of $r_{\rm{max}}$) does not affect much its final chemical
properties (metallicity distribution and [Mg/Fe] vs [Fe/H]). It
acts primarily on enhancing the final
total stellar mass, while it slightly modifies the age distribution of
the stellar population. This latter consequence is mainly seen at low
densities ($\rho_{\rm{c,gas}} \sim 0.015\,\rm{m_{H}/cm^3}$). These points are also illustrated in
Figs.~\ref{fig13_appendix_2} and \ref{fig15_appendix_2}, with
different $c_\star$ and $\epsilon_{SN}$ than in Fig.~\ref{fig1_central_density}.

Conversely, a variation by a factor 10 in the central gas density can
vary the final metallicity by more than a dex, making it the primary
driving parameter.  The cooling time is shorter for larger densities,
enhancing the star formation, and therefore resulting in more luminous
and metal-rich systems.  At high enough central densities, all models,
even the least massive ones, may experience a strong initial burst, as
seen in Fig.~\ref{fig17_appendix_2}.  In the range
$\rho_{\rm{c,gas}}=0.015$ to
$0.037\,\rm{m_{H}/cm^3}$, the star formation history
can be fully modified.  Example is provided in
Fig.~\ref{fig19_appendix_2}, where the central density is varied by a
factor $4$, passing from a low and continuous star formation to an
initial strong burst.

%


\section{The different observed regimes of star formation}\label{regimes_of_starformation}



\begin{figure}
\resizebox{\hsize}{!}{\includegraphics[angle=0]{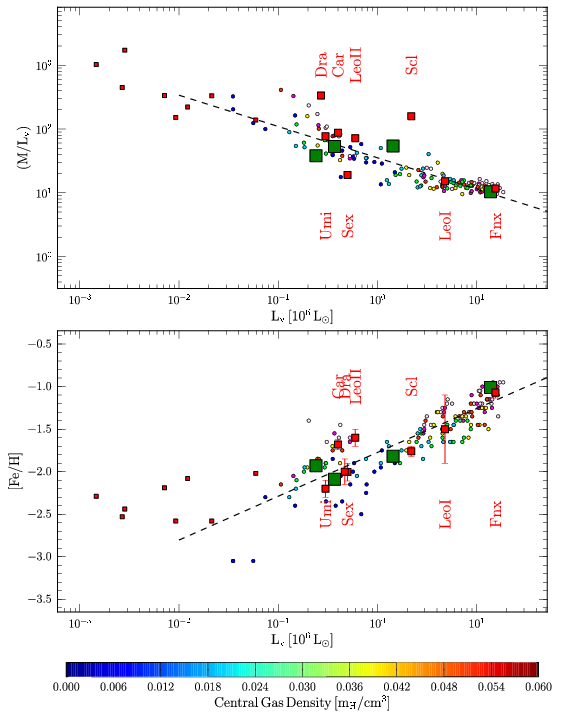}}
\caption{Global relations. The galaxy mass-to-light ratios, $M/L_{\rm{v}}$ (top panel) and 
the mode of their final metallicity distributions (bottom panel),  
versus their $V$-band luminosities  $L_{\rm{V}}$.  
{\rm Each circle stands for one of our 160  models, with  \cstar and $\epsilon_{\rm{SN}}$ restricted to the intervals $0.05-0.1$ and $0.03-0.05$, respectively. 
The central gas density and total mass range is $0.005-0.06\,\rm{m_{H}/cm^3}$ and  $1-9\times 10^8\,\rm{M_\odot}$.}
Colors  code   the initial central gas density.
The red large  squares represent the Local Group classical dSphs, while the smallest red squares stand for the ultra faint dwarfs.  
The dashed lines correspond to our model best fits. The green squares indicate the position of our selected generic
models, from left to right, Carina, Sextans, Sculptor and Fornax.}
\label{fig_regimes_of_starformation_1}
\end{figure}

Fig.\ref{fig_regimes_of_starformation_1} presents in \feh and $M/L$ vs luminosity plots
the full series of simulations, which are globally compatible with the observations.  This represents a
total number of $160$ simulations with $c_{\star}$ in the range
$0.05-0.1$, $\epsilon_{\rm{SN}}$ in $0.03-0.05$, $\rho_{\rm{c,gas}}$
in $0.005-0.06\,\rm{m_{H}/cm^3}$ and $M_{\rm{tot}}$ in $1-9\times
10^8\,\rm{M_\odot}$.  While the reproduction of these global relations
is a necessary step, it is however not sufficient to ensure that the
star formation history of the galaxies are correct.

Among these simulations, we selected a set of 4 models best
reproducing the observations. We limited as much as possible variations in
$\epsilon_{\rm{SN}}$ and $c_{\star}$. 
Our selection was based on
comparison with the observed galaxy metallicity distribution, the
V-band luminosities, the abundance ratios, and the stellar age
distributions.  Observed luminosities were taken from
\citet{walker09}.  The metallicity distributions were retrieved from
\citet{battaglia11} for Sextans, from \citet{battaglia06},
\citet{tolstoy04}, and \citet{helmi06}, for Fornax, Sculptor, Carina,
respectively, with the new CaT calibration of \citet{starkenburg10}.
The abundance ratios were taken from \citet{tolstoy09},
\citet{letarte10}, \citet{shetrone03}, and \citet{koch08}.  The
stellar age distributions were taken from
\citet{smecker-hane96,hurley-keller98} (Carina), \citet{babusiaux05,
  shetrone03, tolstoy03} (Sculptor), \citet{coleman08} (Fornax) , and
\citet{lee03} (Sextans). A summary of the mean observed properties of
the four local group dSphs is given in
Tab.~\ref{tab_regimes_of_starformation_2}.  It also includes the dSph tidal
radii from \citet{irwin95} as well as the velocity dispersions from
\citet{walker09c}.

Fig.\ref{fig_regimes_of_starformation_2} displays our final selection.
These models are shown with green squares in
Fig.~\ref{fig_regimes_of_starformation_1}; they fall close to the
observed $M/L$, \feh and $L_{\rm{V}}$, within a factor 2 for each
quantity.   Table
\ref{tab_regimes_of_starformation_2} provides the model parameters
and outputs. $r_{\rm{t}}$ is the radius encompassing $90\%$ of the
V-band luminosity.  $M_{\rm{gas}}$, $M_{\rm{star}}$, $M_{\rm{halo}}$
and $M/L$ are computed inside $r_{\rm{t}}$.  \feh is calculated as the mode of the galaxy
metallicity.  The stellar and dark matter velocity dispersions,
$\sigma_\star$ and $\sigma_{\rm{DM}}$, are measured along the
\emph{line of sight}, inside the effective radii (half light radius).
We differentiated the dark matter and the stellar velocity dispersions,
because the former best traces the total mass of the system, while the
observations gather the latter.

\subsection{Distinct fates for the SNeII and SNeIa feedbacks}\label{fates_for_feedbacks}

In all systems, the star formation history results from the balance
between cooling and heating. Both SNeII and SNeIa play important
roles, despite their relative different numbers and consequently
different integrated energy feedback (a factor 10 between the two).

\begin{figure*}
\resizebox{\hsize}{!}{\includegraphics[angle=0]{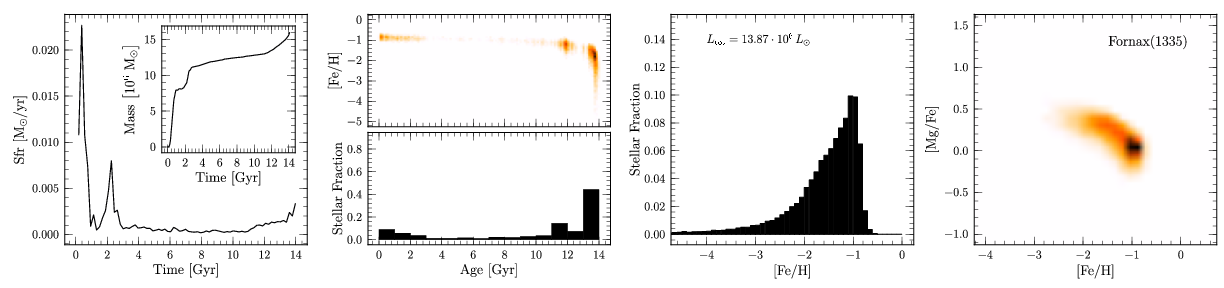}}
\resizebox{\hsize}{!}{\includegraphics[angle=0]{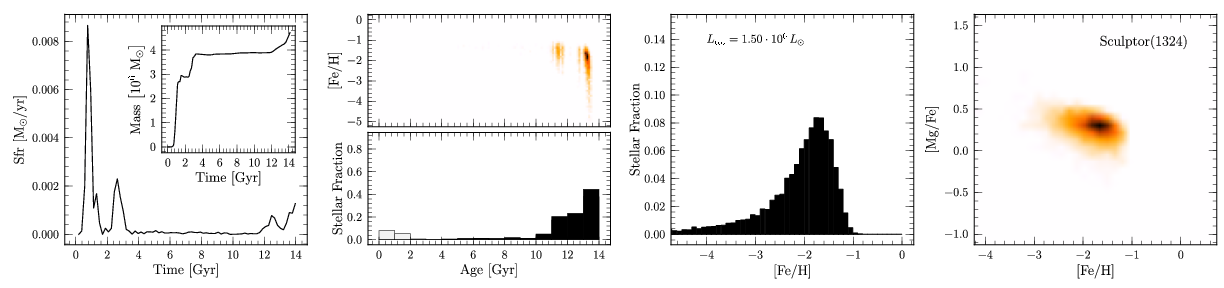}}
\resizebox{\hsize}{!}{\includegraphics[angle=0]{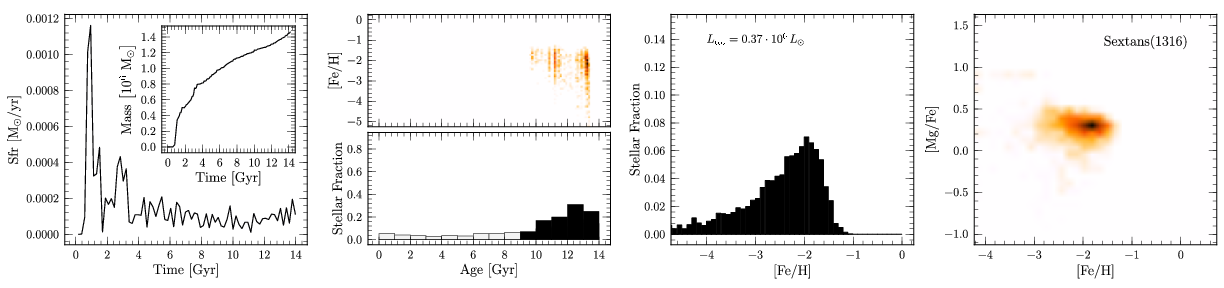}}
\resizebox{\hsize}{!}{\includegraphics[angle=0]{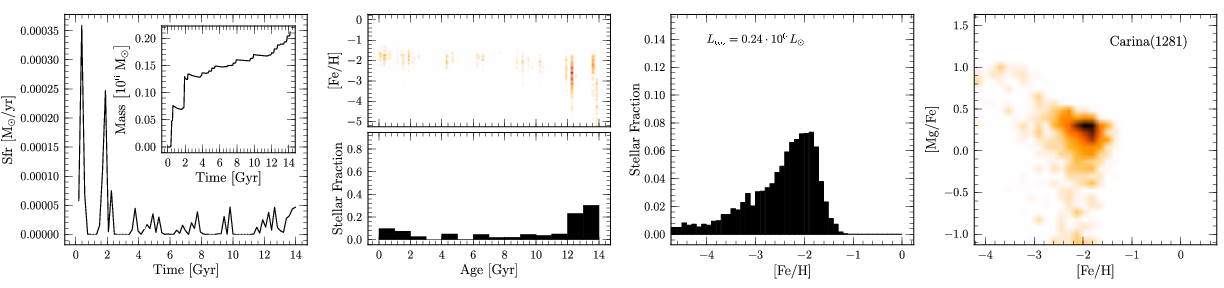}}
\caption{Properties of the four selected models representing Fornax, Sculptor, Sextans and Carina.
From left to right, the plots display, the star formation rate and the evolution of the stellar mass, the 
normalized stellar age distribution, together with the evolution of [Fe/H].
The final [Fe/H] distribution and ratio [Mg/Fe] as a function of [Fe/H].
The grey region in the stellar age distribution of Sculptor and Sextans
correspond to the stellar population which would have been created if we would not have truncated the star formation. These populations are absent from the metallicity distributions and abundance ratios.
The stellar V-band luminosities computed are $14\,\rm{Gyr}$. }
\label{fig_regimes_of_starformation_2}
\end{figure*}

Looking closely at the course of the galaxy evolution, one sees that
star formation is ignited in the galaxy central parts. Massive stars
explode rapidly as SNeII and their injections of energy counteract the
increase of the central gas density, leading to a flat gas profile
within $\sim 1\,\rm{kpc}$.  
The strong and quick energy injection of SNeII explains why the rapid initial rise
in intensity of the star formation rates (see first panels in
Fig.~\ref{fig_regimes_of_starformation_2}) is soon followed by a sharp
decrease: the feedback energy acts against star formation.

Meanwhile, lower mass stars are progressively redistributed along a
steep profile.  Indeed, their velocity dispersion at birth is small,
of the order of $5\,\rm{km/s}$, as inherited from the turbulent gas
motion.  Therefore their ensemble contracts because, being decoupled
from the gas, it is no longer supported by pressure.  The number SNeIa
increases with time until they become the dominant source of heating
in the central regions ($0.2\,\rm{pc}$) preventing the gas to cool and
condense again, despite a very short cooling time.  The dominance of
the heating by SNeIa over the SNeII's one is only possible because the
former are more concentrated than the latter. This causes the few
Gyr-long periods of very low level (or absence) of star formation
after the initial $2-4\,\rm{Gyr}$.  Once the bulk of SNeIa explosions
has passed, star formation is possible again at higher rates.

These phenomena have been explicitly witnessed in a Sculptor-like
simulation in which the energy released by SNeIa was artificially set
to zero. Star formation was then maintained to a substantial level of
$0.005\,\rm{M_{\odot}/yr}$, corresponding to half of the amplitude of
the first burst, where star formation had been quenched in the generic
model.

\subsection{Generic models}\label{generic_models}

Overall, our models mean properties are in excellent agreement with
the observations. This suggests that a sequence of dSphs can be
reproduced by varying very few parameters.  Carina, Sextans, Sculptor
and Fornax form an increasing sequence of total masses and initial
central densities.  All systems display a first $ \sim 2\,\rm{Gyr}$
burst. Depending on the mass, the star formation history can be then
considerably reduced, or even quenched, in the case of small systems.
$\epsilon_{\rm{SN}}$ is fixed to $0.03$; $c_{\star}=0.05$ except for
Carina.  Clearly, as discussed in Section~\ref{cstar}, increasing
$c_{\star}$ favors discontinuous star formation histories.  This small
difference may be the sign of some external factor in the formation of
the galaxy, which was not taken into account in our models.

The sequence in mass and central gas density can be understood by the
following considerations, at fixed $c_{\star}$ ($\sim$ 0.05) and
$\epsilon_{\rm{SN}}$ ($\sim$ 0.03): 

\begin{itemize}

\item At high mass, i.e., above $\sim 3\times 10^8\,\rm{M_\odot}$. The
  evolution of the systems is quite uniform. The cooling always
  dominates the feedback and the virial temperature reaches
  $10^4\,\rm{K}$ around which the radiative cooling is very strong.
  Hence, the star formation is never completely quenched.

\item The diversity of evolutionary paths is larger for lower mass
  systems.  Those who initially form a large quantity of stars compared
  to their total mass, do release a large SNeIa-driven feedback energy,
  which is able to counterbalance their cooling. The level at which  star
  formation is quenched depends on the ratio between the cooling and
  the energy released by SNeIa,  following the process described in Section~\ref{fates_for_feedbacks}.  
  This ratio reflects the level at
  which the galaxy has used its full capacity to form stars during the
  first Gyr, i.e, the fraction of gas transformed into stars, as
  determined by the galaxy mass and initial central density.
   Low mass with low central gas density  systems  have a rather long 
  cooling time ($>1\,\rm{Gyr}$). The SNeII are sufficient to completely quench momentarily
  the star formation, which needs more than $1\,\rm{Gyr}$ to be reignited, and 
  proceeds further episodically.
\end{itemize}

The four models reproducing the observed properties of Carina,
Sextans, Sculptor and Fornax follow these rules.  An additional
feature is that the choice $\epsilon_{\rm{SN}}$
(primarily) and $c_{\star}$ was definitely driven by Fornax. Its
luminosity and mass require a vigorous and continuous star
formation, in turn forcing both parameters to be low.

Just as in \citet{revaz09}, our models retain large masses of gas,
from $10^6$ to $10^7\,\rm{M_\odot}$ and in inverse proportion of the
galaxy stellar mass, i.e., low stellar mass systems have the largest
fraction of gas at $14\,\rm{Gyr}$.  Moreover, both Sextans and
Sculptor models need to have their star formation artificially
stopped, respectively at $\sim 5$ and $\sim 9\,\rm{Gyr}$.
The low
$\epsilon_{\rm{SN}}$ and $c_{\star}$ imposed by Fornax are inefficient
in quenching it either by the means of thermal SNe feedback or even
galactic winds.  This strongly supports stripping of the gas by
external gravitational and/or hydrodynamical interactions (see for
example, \cite{mayer06}), even though the bulk of the galaxy
formation could be driven by their initial conditions.

\begin{table*}
\renewcommand{\footnoterule}{}

  \begin{tabular}{l c c c c c c c | c c c c c c c c c c }
    \hline\hline
    dSphs	& \# &$M_{\rm{tot}}$		& $\rho_{\rm{c,gas}}$ & $r_{\rm{max}}$ & $c_\star$ & $\epsilon_{\rm{SN}}$ & $t_{\rm{trunc}}$ & $L_{\rm{V}}$ & $\langle \rm{[Fe/H]}\rangle$ & $r_{\rm{t}}$ & $\sigma_\star$ & $\sigma_{\rm{DM}}$ & $M_{\rm{gas}}$ & $M_{\rm{stars}}$ & $M_{\rm{halo}}$\\ 
                &       &$10^8\,\rm{M_\odot}$ & $\rm{m_{H}/cm^3}$ & $\rm{kpc}$ & - & - & $\rm{Gyr}$ & $10^6\,\rm{L_\odot}$ &&	$\rm{kpc}$  &$\rm{km/s}$ & $\rm{km/s}$ & $10^7\,\rm{M_\odot}$& $10^7\,\rm{M_\odot}$& $10^7\,\rm{M_\odot}$\\
    \hline\hline
      Fornax  &1335&	7&	0.059&	 7.1	& 0.05&	0.03 & -  &	13.9	&	$-1.01$	& 1.98 & 9.4 & 15.0  & 2.4 & 1.35 & 8.80\\
      Sculptor&1324&	5&	0.029&	 9.6	& 0.05&	0.03 & 9.1&	1.50	&	$-1.75$	& 2.93 & 6.4 & 11.7  & 1.9 & 0.34 & 4.45\\
      Sextans &1316&	3&	0.022&	 8.0	& 0.05&	0.03 & 4.7&	0.37	&	$-2.09$	& 1.58 & 4.2 & 9.7   & 0.5 & 0.07 & 1.04\\
      Carina  &1281&	1&	0.022&	 3.5	& 0.1&	0.03 & -  &	0.24	&	$-1.93$	& 0.76 & 3.1 & 7.2   & 0.2 & 0.02 & 0.63\\
    \hline                                                                 
  \end{tabular}                                                            
  \caption{ Properties of the four best generic models for the Local Group dSphs,  Fornax, Sculptor, Sextans and Carina. Left-ward of the vertical line are the physical input parameters of the models. Right-ward of the same line are the final outputs of the models. They are
calculated  within the radius containing 90\% of the galaxy total light.}
  \label{tab_regimes_of_starformation_1}
\end{table*}

\begin{table}
\renewcommand{\footnoterule}{}

  \begin{tabular}{l c c c c c c c }
    \hline\hline
    dSph	&  $L_{\rm{V}}$ & $\langle \rm{[Fe/H]}\rangle$ & $M/L$ & $r_{\rm{t}}$  & $\sigma$ \\ 
                &  $[10^6\,\rm{L_\odot}]$ &&&	$[\rm{kpc}]$ &$[\rm{km/s}]$ \\
    \hline\hline
      Fornax   &    $14$    & $-1.17$   & 12  & $2.08$& $ 11.7$\\
      Sculptor &    $1.4$   & $-1.96$   & 158 & $1.33$& $  9.2$\\
      Sextans  &   $0.41$   & $-2.26$   & 19  & $3.10$& $  7.9$\\
      Carina   &   $0.24$   & $-1.86$   & 88 &  $0.58$& $  6.6$\\
    \hline                                                                 
  \end{tabular}                                                            
  \caption{Some of the observed properties of the Local Group dSphs Fornax, Sculptor, Sextans, and Carina,
which are used in the present study . The luminosity comes from \citet{walker09}. The mean metallicities were calculated from 
\citet{battaglia11} for Sextans,
from \citep{battaglia06},  \citep{tolstoy04}, and \citep{helmi06},
for Fornax, Sculptor, Carina, respectively, with the new CaT calibration 
of \citet{starkenburg10}.
  The $M/L$ ratios were computed using the masses derived by \citet{walker07, battaglia08, kleyna04}.
  The tidal and core radii are taken from \citet{irwin95} and the velocity dispersions from \citet{walker09c}.}
  \label{tab_regimes_of_starformation_2}
\end{table}

\section{Spatial distribution of the stellar populations}\label{gradients}

The presence of metallicity gradients in dSphs is debated.  Evidence
for  stellar population segregation is reported in Sculptor
\citep{tolstoy04}, Fornax \citep{battaglia06}, and Sextans
\citep{battaglia11}. In these galaxies, the most metal-rich stars are concentrated in the
galaxy central regions while the metal-poor ones are essentially found
at every radius.  Conversely no radial change in the metallicity is
found in LeoI \citep{koch07a,bosler07,gullieuszik09}, LeoII
\citep{koch07b}, and CVnI \citep{ural10}. \cite{kirby11} reported
low metallicity gradients of at most $-0.21$ dex per core
radius in LeoII .

\begin{figure}
\resizebox{\hsize}{!}{\includegraphics[angle=0]{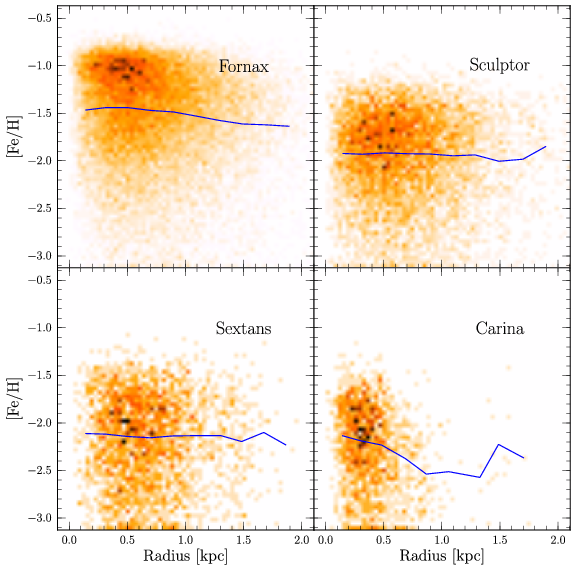}}
\caption{
The radial distribution of the stellar \feh for our four selected generic models. 
Each panel is a 2D  histogram of the number of particles weighted by their mass. 
The blue curve follows the mean value of \feh computed at each radius bin.
}
\label{fig_gradients_1}
\end{figure}

Fig.~\ref{fig_gradients_1} presents the radial \feh density maps for
our four generic models.  They show no evidence for any radial change
in metallicity distribution.  This is a common feature to all our
models. In the following, we examine the gas motion
leading to the final structure of our model galaxies and
consider two of our very high resolution simulations containing
$4'194'304$ particles with an initial total mass of
$3\times10^8\,\rm{M_{\odot}}$ and $9.5\times10^8\,\rm{M_{\odot}}$,
from Table \ref{table_convergence_tests}.

\begin{figure*}
\resizebox{\hsize}{!}{\includegraphics[angle=0]{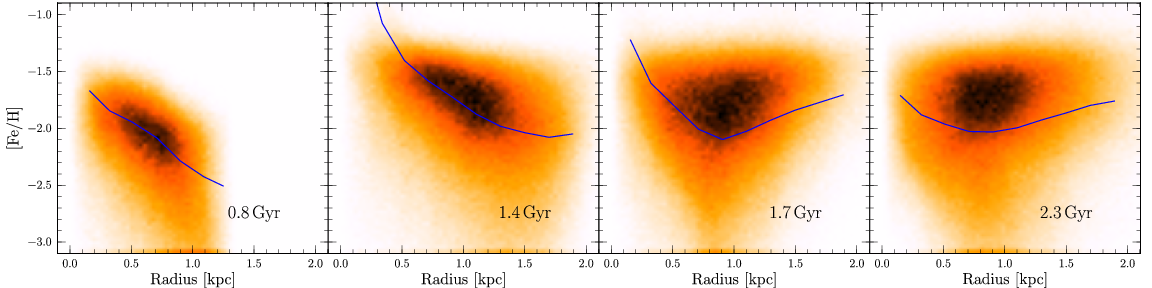}}
\caption{
Evolution of the stellar metallicity gradient with time for a
$3\times10^8\,\rm{M_{\odot}}$ model containing $4'194'304$ particles.
As in Fig.~\ref{fig_gradients_1}, each plot corresponds to a 2D
histogram of the number of particles weighted by their mass.  }
\label{fig_gradients_2}
\end{figure*}

The $3\times10^8\,\rm{M_{\odot}}$ model is characterized by two strong
episodes of star formation (see the top of
Fig.~\ref{fig1_convergence_tests}). The first one occurs between
$0.4$ and $1.6\,\rm{Gyr}$, the second one, slightly weaker, takes
place between $2.1\,\rm{Gyr}$ and $3.2\,\rm{Gyr}$.  The physical
processes described in the following apply to any other model with a
small mass. Fig.\ref{fig_gradients_2}, displays the radial \feh
density maps of the gas during the first period of star formation.
The metallicity gradient seen in the first snapshot at $t=0.8\,\rm{Gyr}$
is the consequence of the first generations of stars forming
preferentially in the galaxy central high density regions, hence
leading to higher chemical enrichment.  This implies that for a short
period of time, stellar evolution occurs on shorter time scales  than
the gas motion.  The gradient persists at $t=1.4\,\rm{Gyr}$,
although being already shallower, but soon after it disappears.

\begin{figure*}
\resizebox{\hsize}{!}{\includegraphics[angle=0]{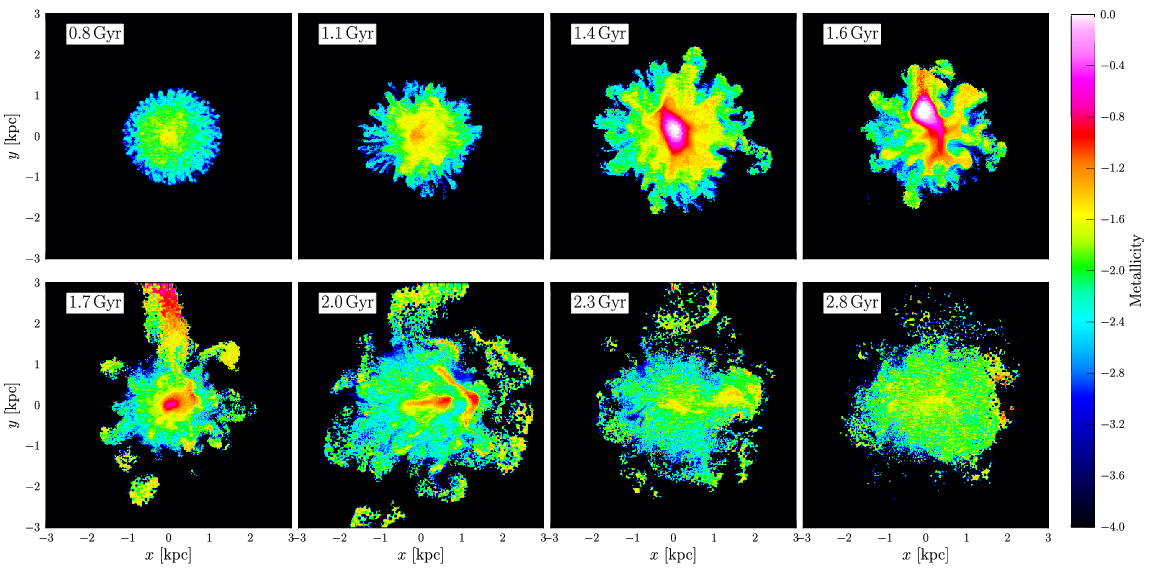}}
\resizebox{\hsize}{!}{\includegraphics[angle=0]{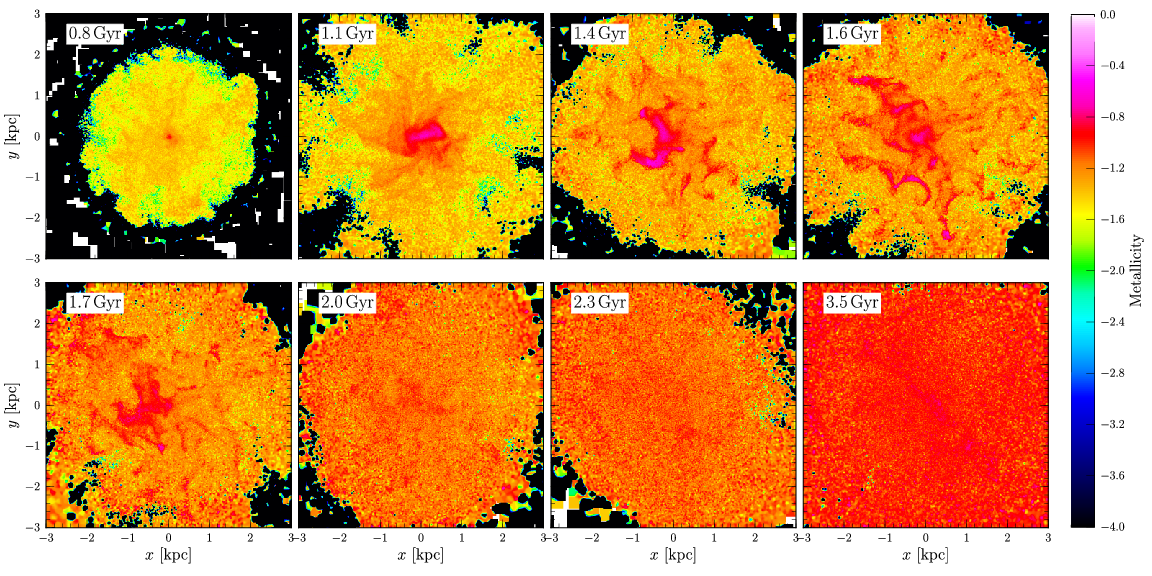}}
\caption{
Evolution of the metallicity with time. Top, model with a total mass of $3\times10^8\,\rm{M_{\odot}}$.
Bottom, the one with $9.5\times10^8\,\rm{M_{\odot}}$.
The maps corresponds to a thin slice of $0.1\,\rm{kpc}$. The colors code $[Fe/H]$, from $-4$ to $0$.
}
\label{fig_gradients_3}
\end{figure*}

Fig.~\ref{fig_gradients_3} presents the physical mechanisms at play.
As seen in the upper panel, the gas accumulate  in the central regions
during the period of intense star formation, not only
metals but also the thermal energy released by the explosions SNeII.
Consequently, the gas is hotter and more tenuous than its
surrounding (see snapshot at $t=1.4\,\rm{Gyr}$) and forms a bubble.
Very strong Archimedes forces act in the vicinity of the center of the
galaxy potential well, making the position of the bubble
unstable. Small motions induced by the local turbulence is sufficient
to offset it and makes it quickly dragged outward.  This phenomenon is
seen twice: first between $1.5$ and $1.7\,\rm{Gyr}$, then between
$1.7$ and $2\,\rm{Gyr}$. The hot and metal-rich gas bubble (seen in
red in Fig.~\ref{fig_gradients_3}) leaves the main body of the galaxy,
reaching regions up to $10\,\rm{kpc}$. Subsequent episodes of star
formation would experience the same scenario, erasing any metallicity
gradient in the gas in less than a Gyr.  We note that the uplift of
hot bubbles by Archimedes forces is a common mechanism in
cooling flow clusters
\citep{revaz08}.

The evolution of the $9.5\times10^8\,\rm{M_{\odot}}$ model is
displayed at the bottom of Fig.~\ref{fig_gradients_3}.  A gradient in
metallicity is visible between $1.1$ and $1.7\,\rm{Gyr}$. The
fundamental difference with the former example is that here stars
efficiently form in a much wider region. Metals and feedback energy
are then injected in a larger volume.  Therefore, turbulent motions
are sufficiently strong to erase the initial gradients.

In summary, our high resolution simulations show that the hot gas motion
have very short timescales of evolution, that are incompatible with the formation 
of stellar metallicity gradient. 
We note that our simulations do not include mixing in the
interstellar medium. This would dilute even more any differential
spatial metallicity distribution. Conversely, a multi-phase and
multi-scale structure of the interstellar medium may prevent the
formation and migration of hot bubbles. This probably deserves future
investigation.  As discussed in Sect.~\ref{esn}, the effect of our
feedback is rather small.  In principle, a stronger feedback would  
increase the turbulence of the ISM and also fight against metallicity
gradients.  

%


\section{The stellar mass fractions in N-body simulation dark matter haloes}\label{stellar_mass_fraction}

The final baryonic fraction f$_{\rm{bf}}$ is an interesting quantity,
since it can be observationally derived in galaxies and is different
from the initial cosmological one.  Within the radius containing 90\%
of the stars, this fraction varies in our models from 25\% to
35\%. Considering the stellar baryonic mass, it ranges from 2\% to
10\%, similarly to the results of \citet{valcke08}, who ran models of
galaxies in isolation as well.

Looking at $M_{\rm{stars}}/M_{\rm{halo}}$, \citet{sawala11} compared
the predictions of a series of N-body models of galaxies with the
expected ratios calculated from the Sloan Digital Sky Survey (SDSS)
for the stellar part and from the MilleniumII simulation for the pure
dark matter content \citep{guo10}. The dependence of
$M_{\rm{stars}}/M_{\rm{halo}}$ on $M_{\rm{halo}}$ was originally
performed for stellar masses between $10^{8.3}\rm{M_\odot}$ and
$10^{11.8}\rm{M_\odot}$. It was then extrapolated from
$10^{8.3}\rm{M_\odot}$ down to $10^6\rm{M_\odot}$. Doing so,
\citet{sawala11} found the fraction of stellar baryons in dark halo
masses below $10^{10}\rm{M_\odot}$ to be lower in the SDSS/MilleniumII
predictions than in the galaxy models. Their interpretation was that
galaxy models may  produce too many stars. This conclusion
deserves further consideration.

\begin{figure}
\resizebox{\hsize}{!}{\includegraphics[angle=0]{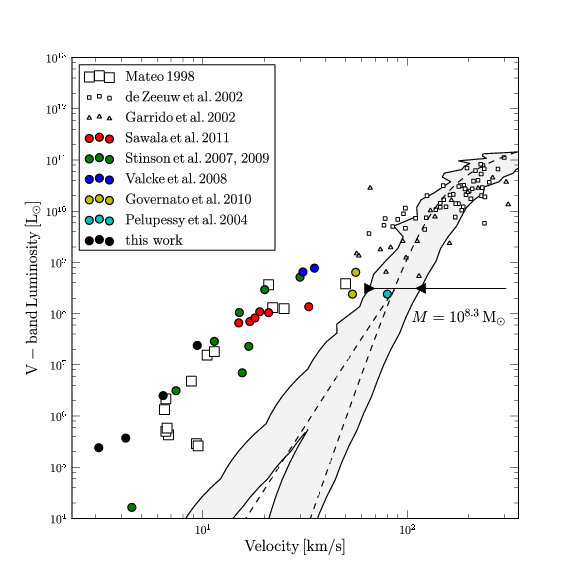}}
\caption{ The relation between the galaxy velocity dispersions or maximal velocities
and their V-band luminosities. Filled circles correspond to the
results of this work and those of N-body simulations. The observations
of low-mass galaxies are displayed with open symbols. The gray area
indicates how the \citet{guo10}'s relation between stellar and dark
halo masses, further extrapolated by
\cite{sawala11}, translates into luminosities and velocity dispersions
for a minimum and maximum value of the NFW concentration
parameter. Below a stellar mass of $10^{8.3}\rm{M_\odot}$, two 
faint end slopes of stellar mass functions are considered (see text),
splitting in two different branches the grey area.  The dashed black
lines are obtained when considering the Aquarius project's relation
between the maximal velocity and halo mass
\citep{springel08}. The minimum stellar mass ($10^{8.3}\rm{M_\odot}$ )
for the original \citet{guo10}'s fit and its correspondence in terms
of range of velocities is indicated with black arrows. }
\label{fig_mssigma_1}
\end{figure}

Figure \ref{fig_mssigma_1} moves the predictions of \citet{sawala11}
into observable quantities.  \cite{guo10}'s relation between galaxy
stellar and halo masses was turned into V-band luminosities and
velocity dispersions. Following \citet{sawala11}, we consider two
different slopes for the faint-end of the stellar mass function,
$-1.15$ \citep{li09} and $-1.58$ \citep{baldry08}. Turning stellar
masses into luminosities was done with a reasonable stellar
$M_{\rm{stars}}/\rm{L_V}$ ratio of 0.75 \citep{flynn06}. 
As the observed V-band luminosities vary over several dex, uncertainties on
 $M_{\rm{stars}}/\rm{L_V}$ impact hardly  the results.
The correspondence between the halo masses and their velocities was
computed in two ways: first, we used the relation between the maximal
velocity and halo mass measured in the Aquarius project
\citep{springel08}. Second, we assumed an NFW profile and solved numerically the
Jeans equations in spherical coordinates following Eq.~\ref{siga_i}, and considering  extreme
values, 5 and 30, for the profile concentration \citep{maccio07}.
The velocity dispersions extracted from the inner $1\,\rm{kpc}$ are computed
by properly taking into account projection effects.

The comparison with the observations - dwarf galaxies \citep{mateo98},
together with irregular, elliptical and spiral galaxies from
\citep{garrido02} and the SAURON sample \citep{dezeeuw02} - and a
range of different N-body simulations
\citep{sawala11,stinson07,stinson09,valcke08,governato10,pelupessy04}
call for a few comments: Noteworthily, the agreement between all
models of small mass systems on the one hand and their consistency
with the observations on the other hand is quite remarkable, despite
considerable variations in the assumptions, including the initial
conditions, from cosmological to isolation.  Where stellar masses were
actually measured in the SDSS, the consistency of \cite{guo10}'s
relation with the observations is very good indeed. Conversely, at
lower masses, the discrepancy with \cite{sawala11}'s extrapolation is
very large.  Whilst the observed velocities are derived from baryonic
matter and could undermine the dark matter halo ones by a factor 2
(see Table \ref {tab_regimes_of_starformation_1}), this is
insufficient to fill in the gap between the SDSS/MilleniumII relation
and the observations.  Similarly, varying the stellar
$M_{\rm{stars}}$/ $\rm{L_V}$ even by a factor 10 does help either. 

The velocity dispersion of the stellar component  depends also on
its density profile, which could well be different from
the dark matter's one. Therefore we performed some tests assuming
different profiles for the stars and the dark haloes.  For the stellar density profile, we
took a truncated plummer sphere \citep[e.g.,][]{battaglia06,battaglia08}, while the dark
matter density profile was kept to the NFW form. We calculated
the inner 1kpc
velocity dispersions of both components; they  agree to within a few percents. This is much
lower than the mean factors observed in Fig.~\ref{fig_mssigma_1}.

In summary, the relation between $M_{\rm{stars}}$ derived from observed
luminosities, color, spectroscopy etc., and $M_{\rm{halo}}$ from
pure dark matter simulations should likely be revisited. The assumption
under which the number density of galaxies and dark matter haloes
match each other \citep{guo10} lead to inconsistency with the
observations at low stellar masses.

%
%



\section{Conclusions}\label{conclusion}

We described our new parallel Nbody/Tree-SPH code,
\texttt{GEAR}. Starting from the public version of \texttt{Gadget-2},
we included the physics of baryons, i.e., metal-dependent gas cooling,
star formation recipes, thermal feedback due to Type Ia and Type II
supernovae explosions, and chemical evolution. \texttt{GEAR} has been
conceived to sustain  high spatial resolution together with detailed
chemical diagnostics, and to follow the galaxy evolution over a full
Hubble time, either in isolation or in a cosmological context. We
qualified his performances with the case of dSph galaxies.

$\rhd$ \texttt{GEAR} conserves the total energy of the systems to
better than $5\%$ over $14\,\rm{Gyr}$. Moreover, the code proved
excellent convergence of the results with numerical resolution.  We
could show that for small galactic systems, such as dSphs, $2^{16}$
particles offer a good trade-off between CPU time and resolution,
catching the essence of the physics at play and allowing secure
predictions.

$\rhd$ Varying the initial random number seed, we estimated that our
intrinsic errors were below 30\% on the model stellar masses, the
V-band luminosities and mode metallicities.

$\rhd$ Hundreds of simulations were performed in order to understand and
quantify the effect of the free parameters, such as the star
formation parameter ($c_\star$), the star formation density threshold
($\rho_{\rm{sfr}}$), the supernova efficiency ($\epsilon_{\rm{SN}}$),
the number of stellar particles formed from each gas particle
($N_\star$), the adiabatic period ($t_{\rm{ad}}$) attached to the
thermal feedback,and the number of particles in a softening radius
($N_{\rm{ngb}}$). The most sensitive parameter is definitely  
$\epsilon_{\rm{SN}}$, with an acceptable range of values between
$0.03$ and $0.05$ to reproduce the sequence of dSphs. 
These low values further imply  that strong winds are incompatible
with their observed metallicities.


$\rhd$ We ran a $512^3$ $\Lambda$CDM cosmological simulation of
pure dark matter in order to study the profiles of haloes with masses
between $10^8$ and $10^9\,\rm{M_\odot}$, that are typical of dSphs.
The expansion of the Universe leads to the formation of stable systems
after $z=6$.  In turn, this justifies models of dSphs in a static
Euclidean space, where the expansion of the universe is neglected.
The physics of baryons that depends on the density in physical
coordinates is correct.  We showed that haloes experiencing only minor
mergers since z = 6 do exist.  Whilst at fixed mass, the densities of
these haloes are very similar, still they exhibit a small dispersion
(factor 3 to 4).

$\rhd$ We confirmed that the total initial mass plays a primordial
role in the evolution of the galaxies. However, we demonstrated that
the initial central gas density is as crucial. Changing the mass by a
factor 10 translates into a few tenth of a dex in the final mean
metallicity, while a factor 10 in central density can increase [Fe/H]
by more than 1 dex. We showed that the Local Group classical dSph
could have their properties reproduced as a sequence of
mass and density.

$\rhd$ Differences in mass and initial gas central densities seen in 
the $\Lambda$CDM simulation lead to a
variety of star formation histories explaining the diversity of the
chemical properties observed in dSphs.  In massive and dense systems,
the cooling dominates the feedback, and stars are form continuously,
leading to luminous and metal-rich galaxies.  At lower masses, the
variety of star formation results from the subtle balance between the
cooling and supernovae feedback of both SNeIa and SNeII. Gas is still
present in our model galaxies with masses of about $10^6$ to
$10^7\,\rm{M_\odot}$ after $14\,\rm{Gyr}$.  This strongly supports
the existence external processes like tidal or ram pressure stripping,
which we did not include in this study.

$\rhd$ Because SNeIa and SNeII have different spatial distributions, they act
in distinct ways in the heating of the systems. In particular, we
showed that the few Gyr-long quiescent periods of star formation are
due to the explosion of the centrally concentrated SNeIa.

$\rhd$ We investigated the relationship between the stellar mass of
galaxies and their parent dark matter haloes. The agreement between
all models of small mass systems on the one hand and with the
observations on the other hand is quite remarkable, despite
considerable variations in the assumptions, including the initial
conditions, from cosmological to isolation. They are consistent with
the observations.

$\rhd$ Despite the fact that the new stars are preferentially formed
in the galaxy central regions, none of our models displays any
segregation in stellar population, such as the metal-rich being more
centrally concentrated than the less metallic ones, as reported in
some observations. We showed how turbulence in massive systems and
uplifted hot metal-rich bubbles in less massive ones do erase the
initial metallicity gradients seen in the gas at ages younger than $2\,\rm{Gyr}$.
Further investigations are definitely
required to explain the metallicity gradients reported in some
of the classical dSphs.


\begin{acknowledgements}
P.J. and Y.R. were stimulated by the vivid escort of the DART collaboration.
They are grateful to Volker Springel for making \texttt{Gadget-2}
publicly available and for providing a complete and clear documentation.
This work was supported by the Swiss National Science Foundation.
The authors thank the International Space Science Institute (ISSI) at Bern
for their funding of the team ``Defining the full life-cycle of dwarf galaxy evolution: the
Local Universe as a template''.
\end{acknowledgements}


\begin{appendix}


\section{Metals ejection and energy conservation}\label{appendix1}

In this appendix, we will describe precisely how the velocities of the neighboring
stellar particles $i$ are modified in order to improve the conservation of energy .

The mass received by the particle $j$ is~:
        \begin{equation}
        m'_{j} = m_{j} + w_{ij}\,M_{\rm{e}}.
        \end{equation}
where $M_{\rm{e}}$ is the mass ejected by the stellar particle $i$.
Similarly, the mass of element $k$ is~:
        \begin{equation}
        m'_{j}(k) = m_{j}(k) + w_{ij}\,M_{\rm{e}}(k).
        \end{equation}
Here $w_{ij}$ is~: 
        \begin{equation}
        w_{ij} = \frac{m_j W(r_{ij},h_i)}{\rho_i}.
        \end{equation}

The final mass the the stellar particle $i$ is~:
        \begin{equation}
        m'_{i} = m_{i} - M_{\rm{e}}.
        \end{equation}

As the mass of particles changes, in order to conserve the total energy, 
it is necessary to modify the velocities of the particles. We neglect to
correct the change in potential energy.
Before the mass redistribution, the energy of the particles involved is~:
        \begin{equation}
        E = \sum_j \frac{1}{2} m_j v^2_j +  \frac{1}{2} m_i v^2_i,
        \end{equation}
and after, it is~:
        \begin{equation}
        E' = \sum_j \frac{1}{2} m_j v'^2_j + \sum_j \frac{1}{2} w_{ij}\,M_{\rm{e}} v'^2_j +  \frac{1}{2} m_i v^2_i - \frac{1}{2} M_{\rm{e}} v^2_i,
        \end{equation}
where we have assumed that the particle $i$ does not change its velocity.
If $v'^2_j$ are chosen such that~:
        \begin{equation}
               \frac{1}{2} m_j v'^2_j +        \frac{1}{2} w_{ij}\,M_{\rm{e}} v'^2_j =          \frac{1}{2} m_j v^2_j + \frac{1}{2} w_{ij}\,M_{\rm{e}} v^2_i,
        \label{a1}
        \end{equation}
the the summing over $j$ gives~:
        \begin{equation}
        \sum_j \frac{1}{2} m_j v'^2_j + \sum_j \frac{1}{2} w_{ij}\,M_{\rm{e}} v'^2_j =   \sum_j \frac{1}{2} m_j v^2_j + \frac{1}{2} M_{\rm{e}} v^2_i,
        \end{equation}
which implies that~:
        \begin{equation}
        E' = E.
        \end{equation}
The new velocity for each particle is deduced from Eq.~\ref{a1}~:
        \begin{equation}
        v'^2_j = \frac{m_j v^2_j + w_{ij}\,M_{\rm{e}} v^2_i}{m_j +  w_{ij}\,M_{\rm{e}}} = \frac{m_j}{m'_j} v^2_j +    \frac{w_{ij}\,M_{\rm{e}}}{m'_j} v^2_i .
        \end{equation}
The modification of the square of the velocity is due to two terms. The velocity is first decreased, due to the increase of the mass of the particle.
The second term corresponds to the decrease of energy of particle $i$ due to its decrease in mass.
As in practice, $w_{ij}\,M_{\rm{e}}$ is much smaller that $m_j$, the change in velocity is small.
Only the norm of the velocity is affected during the modification of the velocities.

\begin{figure}
\resizebox{\hsize}{!}{\includegraphics[angle=0]{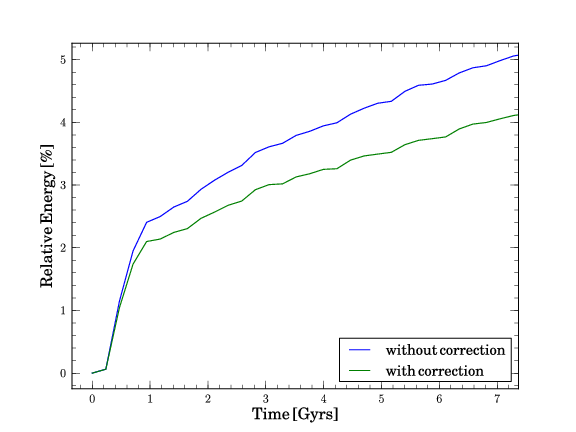}}
\caption{Evolution of the relative energy as a function of time. The green curve corresponds to the model presented in Fig.~\ref{fig1_energy_conservation}. 
It includes the velocity correction while the blue one is similar but uncorrected.}
\label{fig1_appendix_1}
\end{figure}

The effect of this correction on the total energy is estimated by running the
same simulation used in Fig.~\ref{fig1_energy_conservation}, but with the correction switched off.
The comparison of the evolution of the relative energy is given in Fig.~\ref{fig1_appendix_1}. 
The improvement is of about $20\%$ over $7\,\rm{Gyrs}$, while its increase in CPU time is insignificant, we used it.

We didn't observed any improvement in the linear momentum conservation. This is related
to the poor conservation of the  \texttt{treecode} method that does not fulfill Newton's third law and dominates
the error.

\section{Free parameters}\label{appendix2}

\begin{figure*}
\resizebox{\hsize}{!}{\includegraphics[angle=0]{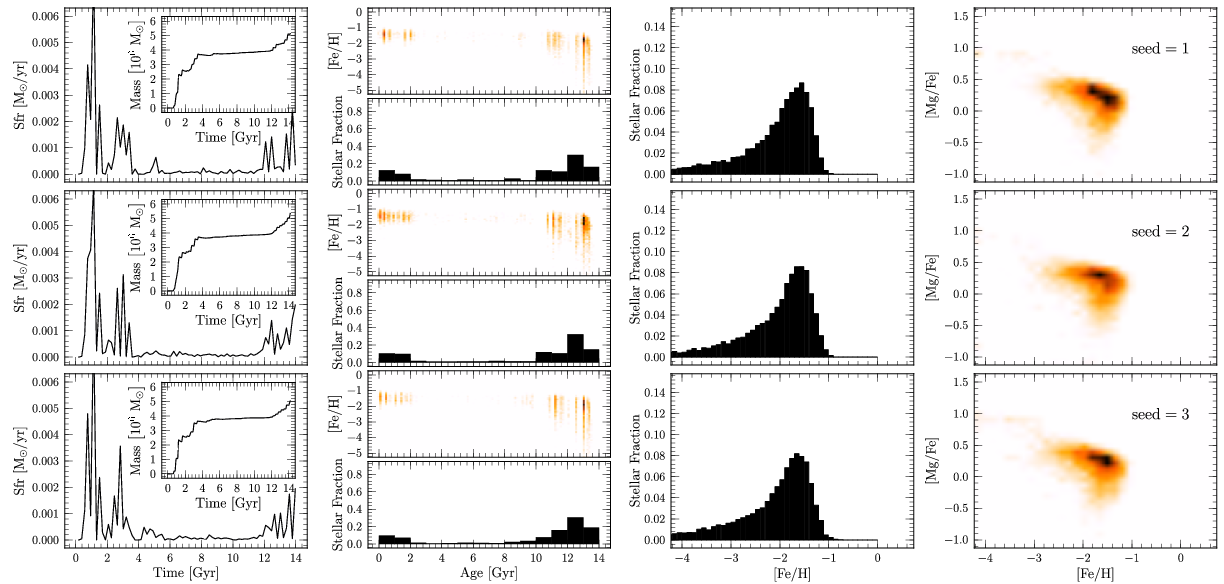}}
\caption{Effect of the random number seed. 
The parameters for the initial conditions are
$M_{\rm{tot}}=8\times 10^8\,\rm{M_\odot}$, $\rho_{\rm{c,gas}}=0.029\,\rm{m_{H}/cm^3}$, $r_{\rm{max}}=8\,\rm{kpc}$.
The parameters for the star formation and supernova feedback are $c_\star=0.1$ and $\epsilon_{\rm{SN}}=0.05$. 
}
\label{fig0_appendix_2}
\end{figure*}

\begin{figure*}
\resizebox{\hsize}{!}{\includegraphics[angle=0]{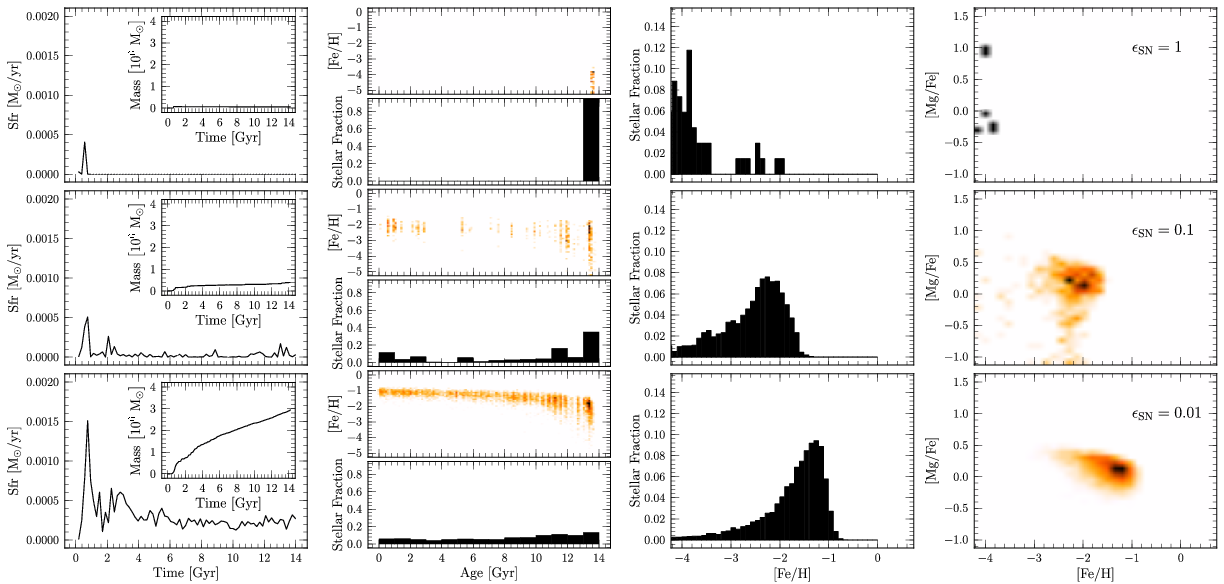}}
\caption{Effect of supernova efficiency $\epsilon_{\rm{SN}}$.
The parameters for the initial conditions are
$M_{\rm{tot}}=3\times 10^8\,\rm{M_\odot}$, $\rho_{\rm{c,gas}}=0.022\,\rm{m_{H}/cm^3}$, $r_{\rm{max}}=8\,\rm{kpc}$.
The parameter for the star formation is $c_\star=0.03$. 
}
\label{fig1_appendix_2}
\end{figure*}

\begin{figure*}
\resizebox{\hsize}{!}{\includegraphics[angle=0]{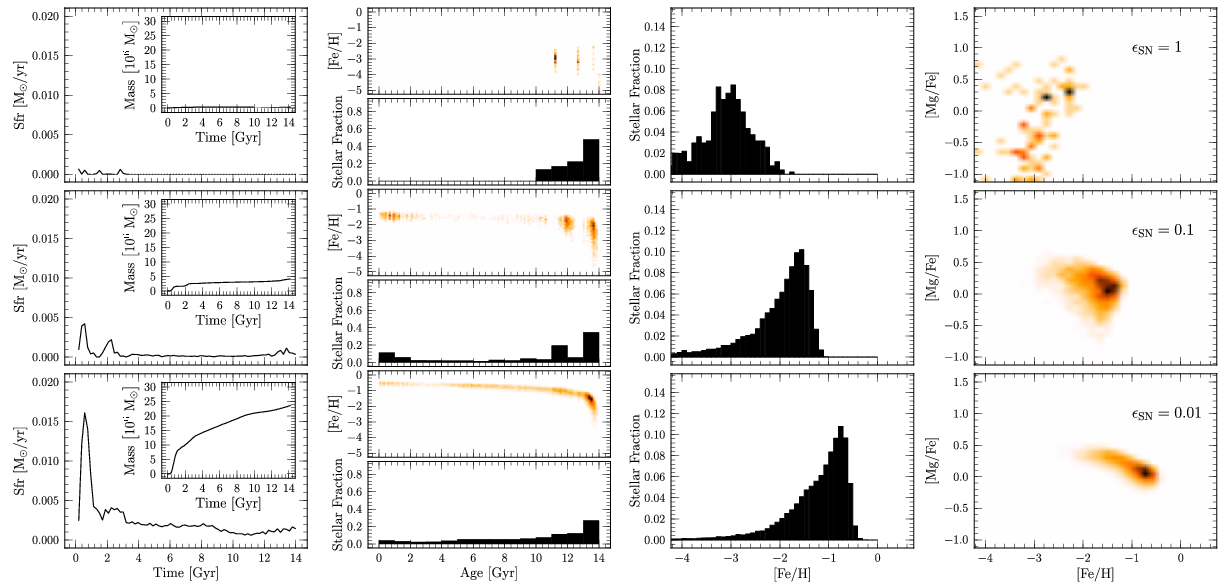}}
\caption{Effect of supernova efficiency $\epsilon_{\rm{SN}}$.
The parameters for the initial conditions are
$M_{\rm{tot}}=6\times 10^8\,\rm{M_\odot}$, $\rho_{\rm{c,gas}}=0.044\,\rm{m_{H}/cm^3}$, $r_{\rm{max}}=8\,\rm{kpc}$.
The parameter for the star formation is $c_\star=0.03$. }
\label{fig2_appendix_2}
\end{figure*}

\begin{figure*}
\resizebox{\hsize}{!}{\includegraphics[angle=0]{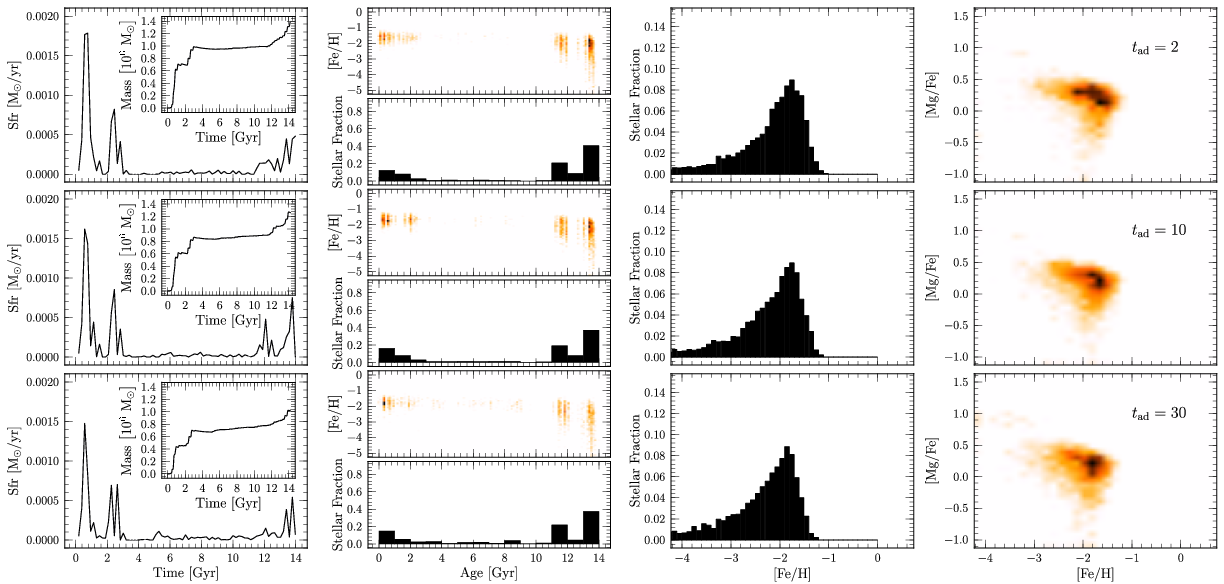}}
\caption{Effect of adiabatic time $t_{\rm{ad}}$.
The parameters for the initial conditions are
$M_{\rm{tot}}=3.5\times 10^8\,\rm{M_\odot}$, $\rho_{\rm{c,gas}}=0.025\,\rm{m_{H}/cm^3}$, $r_{\rm{max}}=8\,\rm{kpc}$.
The parameters for the star formation and supernova feedback are $c_\star=0.05$ and $\epsilon_{\rm{SN}}=0.05$. 
}
\label{fig3_appendix_2}
\end{figure*}

\begin{figure*}
\resizebox{\hsize}{!}{\includegraphics[angle=0]{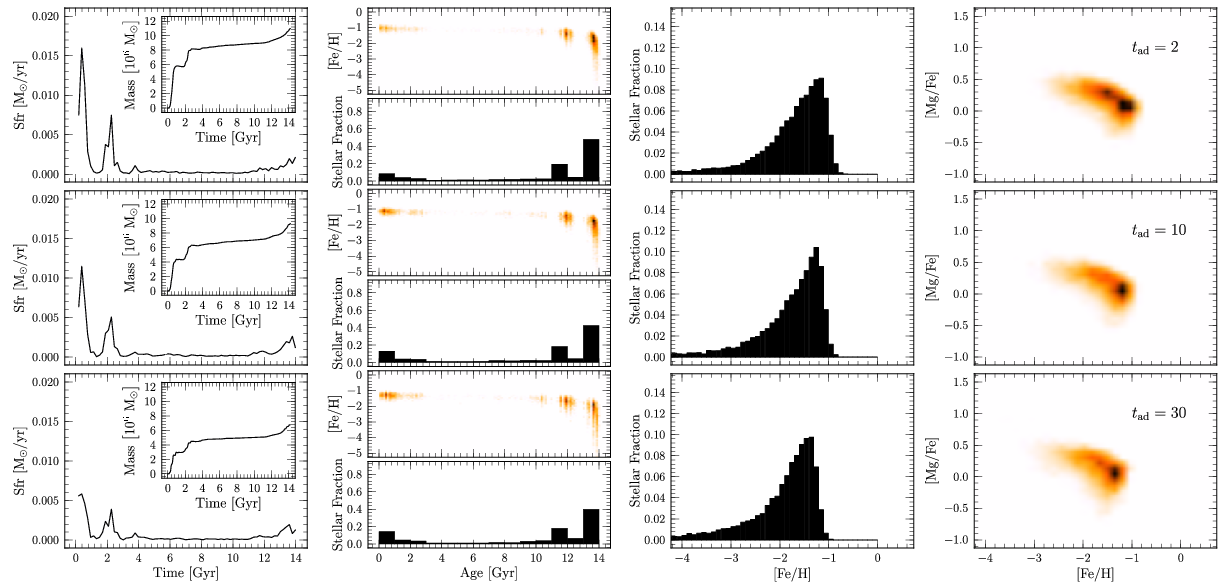}}
\caption{Effect of adiabatic time $t_{\rm{ad}}$.
The parameters for the initial conditions are
$M_{\rm{tot}}=7\times 10^8\,\rm{M_\odot}$, $\rho_{\rm{c,gas}}=0.053\,\rm{m_{H}/cm^3}$, $r_{\rm{max}}=8\,\rm{kpc}$.
The parameters for the star formation and supernova feedback are $c_\star=0.05$ and $\epsilon_{\rm{SN}}=0.05$. 
}
\label{fig4_appendix_2}
\end{figure*}

\begin{figure*}
\resizebox{\hsize}{!}{\includegraphics[angle=0]{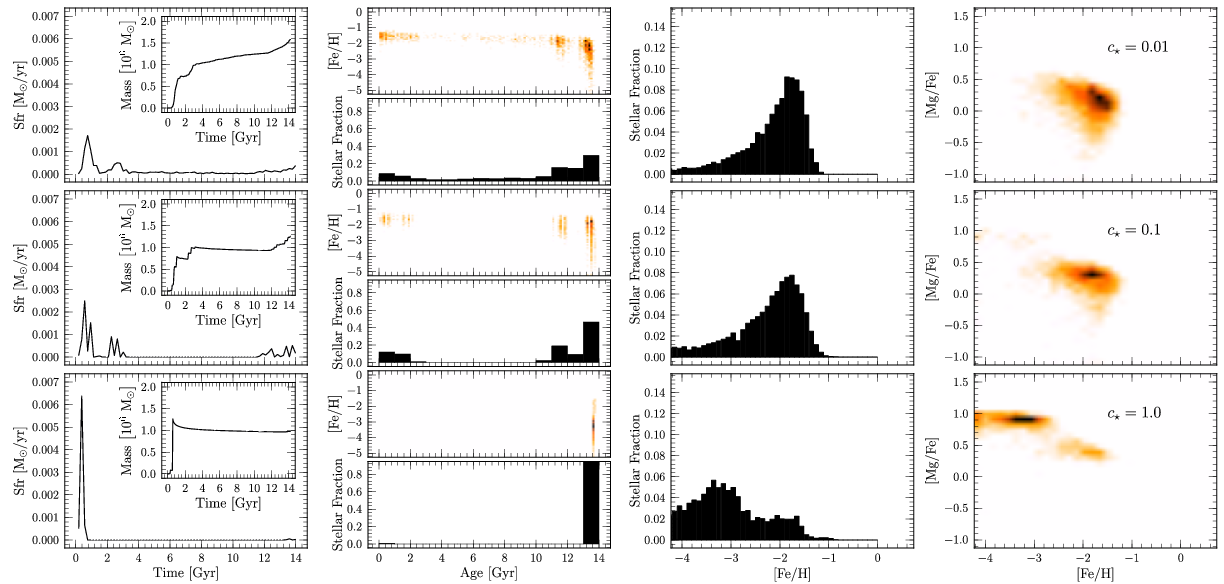}}
\caption{Effect of the star formation parameter $c_{\star}$.
The parameters for the initial conditions are
$M_{\rm{tot}}=3.5\times 10^8\,\rm{M_\odot}$, $\rho_{\rm{c,gas}}=0.025\,\rm{m_{H}/cm^3}$, $r_{\rm{max}}=8\,\rm{kpc}$.
The parameters for the supernova feedback is $\epsilon_{\rm{SN}}=0.05$.
}
\label{fig5_appendix_2}
\end{figure*}

\begin{figure*}
\resizebox{\hsize}{!}{\includegraphics[angle=0]{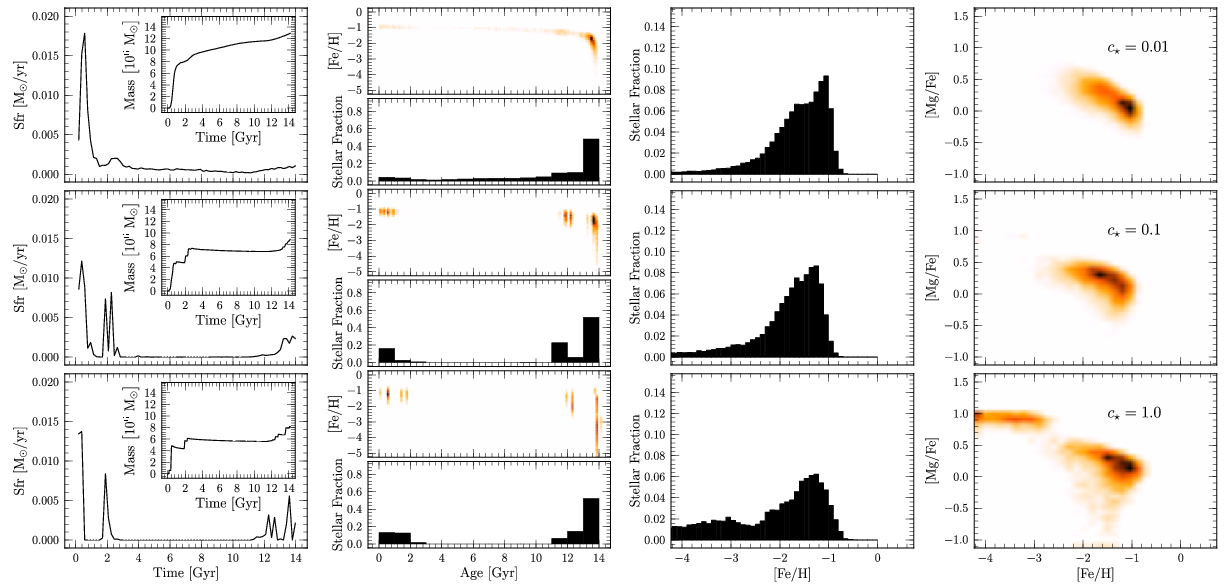}}
\caption{Effect of the star formation parameter $c_{\star}$.
The parameters for the initial conditions are
$M_{\rm{tot}}=7\times 10^8\,\rm{M_\odot}$, $\rho_{\rm{c,gas}}=0.053\,\rm{m_{H}/cm^3}$, $r_{\rm{max}}=8\,\rm{kpc}$.
The parameters for the supernova feedback is $\epsilon_{\rm{SN}}=0.05$.
}
\label{fig6_appendix_2}
\end{figure*}

\begin{figure*}
\resizebox{\hsize}{!}{\includegraphics[angle=0]{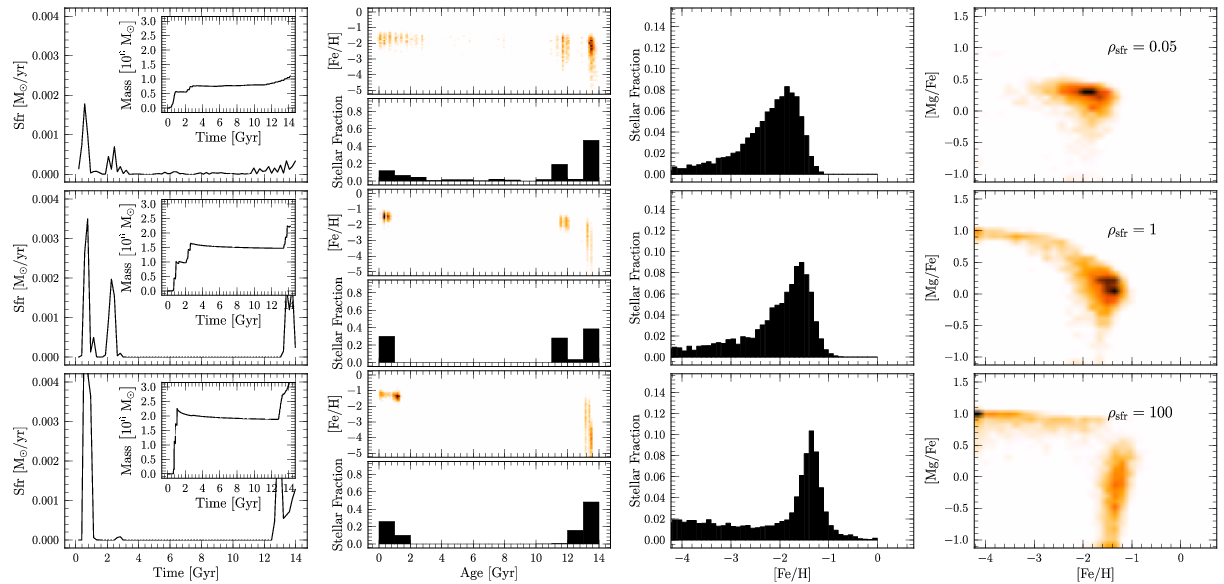}}
\caption{Effect of the star formation parameter $\rho_{\rm{sfr}}$.
The parameters for the initial conditions are
$M_{\rm{tot}}=3.5\times 10^8\,\rm{M_\odot}$, $\rho_{\rm{c,gas}}=0.025\,\rm{m_{H}/cm^3}$, $r_{\rm{max}}=8\,\rm{kpc}$.
The parameters for the star formation and supernova feedback are $c_\star=0.05$ and $\epsilon_{\rm{SN}}=0.05$.
}
\label{fig7_appendix_2}
\end{figure*}

\begin{figure*}
\resizebox{\hsize}{!}{\includegraphics[angle=0]{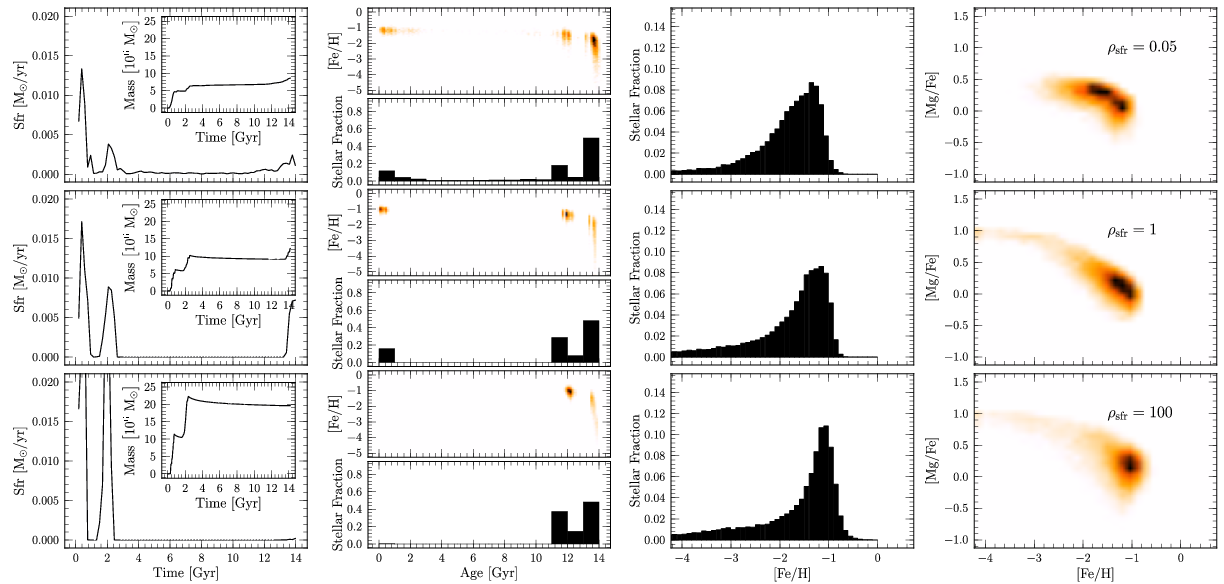}}
\caption{Effect of the star formation parameter $\rho_{\rm{sfr}}$.
The parameters for the initial conditions are
$M_{\rm{tot}}=7\times 10^8\,\rm{M_\odot}$, $\rho_{\rm{c,gas}}=0.053\,\rm{m_{H}/cm^3}$, $r_{\rm{max}}=8\,\rm{kpc}$.
The parameters for the star formation and supernova feedback are $c_\star=0.05$ and $\epsilon_{\rm{SN}}=0.05$.
}
\label{fig8_appendix_2}
\end{figure*}

\begin{figure*}
\resizebox{\hsize}{!}{\includegraphics[angle=0]{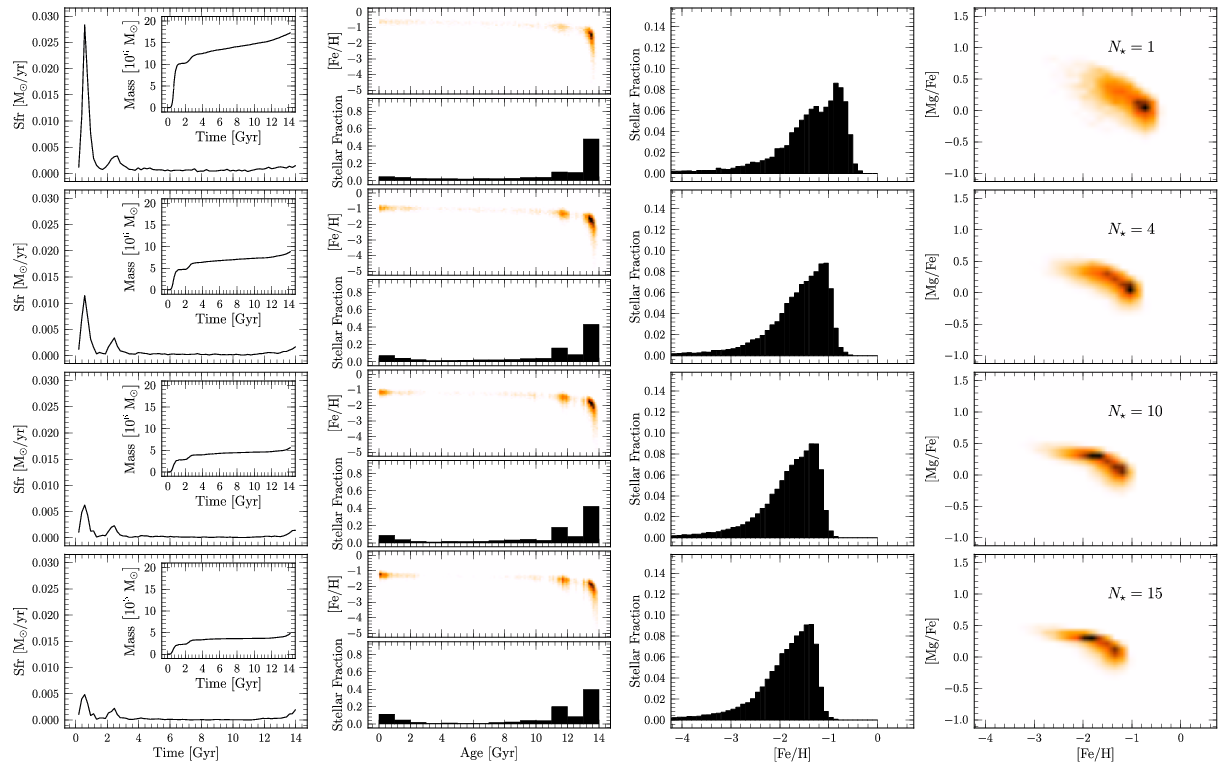}}
\caption{Effect of varying the number of stars formed per gas particle $N_\star$.
The parameters for the initial conditions are
$M_{\rm{tot}}=5\times 10^8\,\rm{M_\odot}$, $\rho_{\rm{c,gas}}=0.037\,\rm{m_{H}/cm^3}$, $r_{\rm{max}}=8\,\rm{kpc}$.
The parameters for the star formation and supernova feedback are $c_\star=0.025$ and $\epsilon_{\rm{SN}}=0.02$.}
\label{fig9_appendix_2}
\end{figure*}

\begin{figure*}
\resizebox{\hsize}{!}{\includegraphics[angle=0]{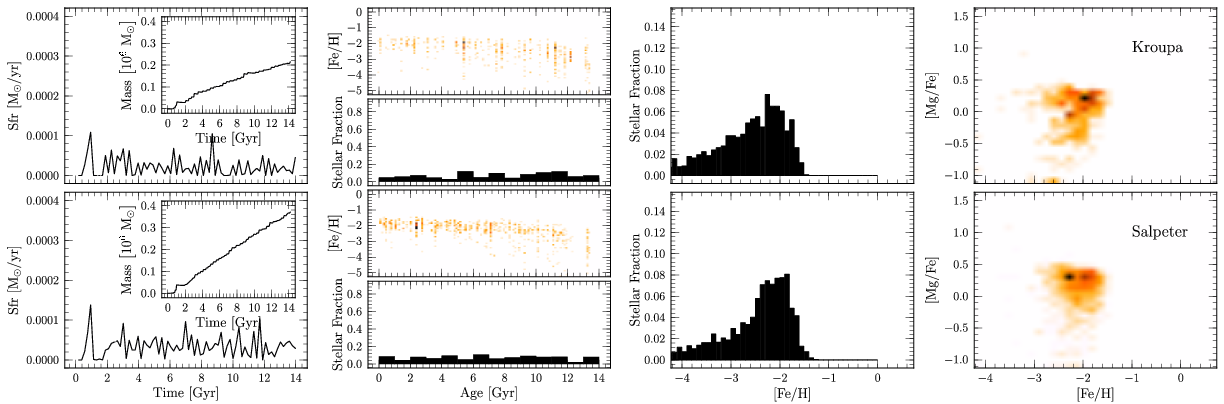}}
\caption{Effect of varying the IMF.
The parameters for the initial conditions are
$M_{\rm{tot}}=2\times 10^8\,\rm{M_\odot}$, $\rho_{\rm{c,gas}}=0.015\,\rm{m_{H}/cm^3}$, $r_{\rm{max}}=8\,\rm{kpc}$.
The parameters for the star formation and supernova feedback are $c_\star=0.05$ and $\epsilon_{\rm{SN}}=0.03$.
}
\label{fig10_appendix_2}
\end{figure*}

\begin{figure*}
\resizebox{\hsize}{!}{\includegraphics[angle=0]{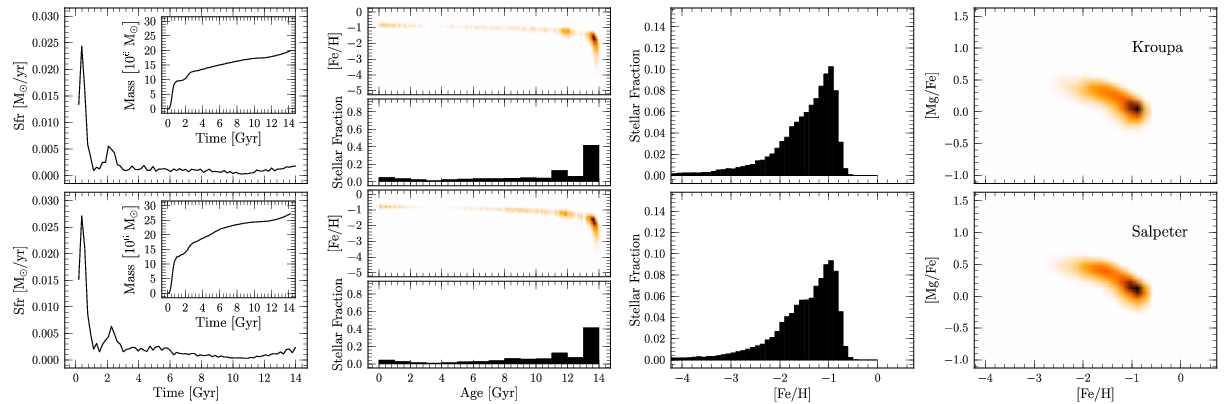}}
\caption{Effect of varying the IMF.
The parameters for the initial conditions are
$M_{\rm{tot}}=8\times 10^8\,\rm{M_\odot}$, $\rho_{\rm{c,gas}}=0.059\,\rm{m_{H}/cm^3}$, $r_{\rm{max}}=8\,\rm{kpc}$.
The parameters for the star formation and supernova feedback are $c_\star=0.05$ and $\epsilon_{\rm{SN}}=0.03$.
}
\label{fig11_appendix_2}
\end{figure*}

\begin{figure*}
\resizebox{\hsize}{!}{\includegraphics[angle=0]{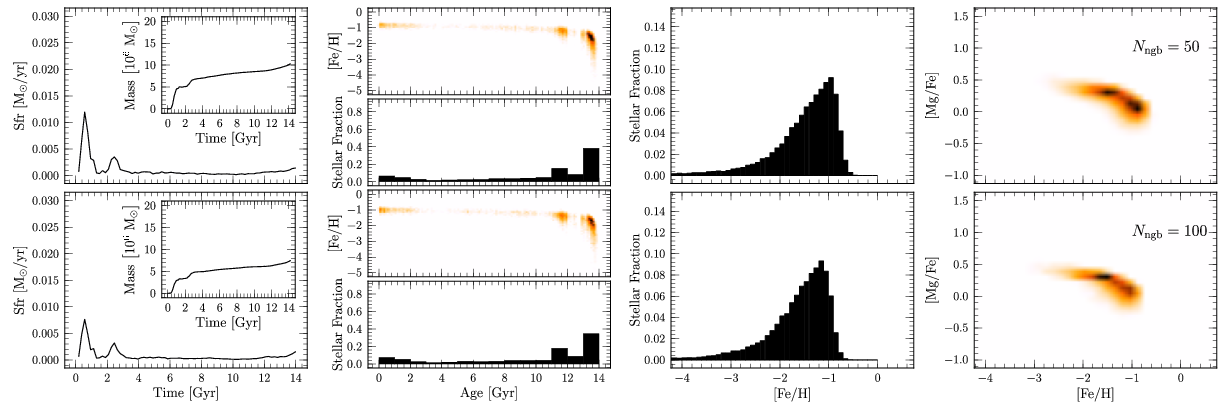}}
\caption{Effect of varying the  number of neighbors, $N_{\rm{ngb}}$.
The parameters for the initial conditions are
$M_{\rm{tot}}=4.5\times 10^8\,\rm{M_\odot}$, $\rho_{\rm{c,gas}}=0.033\,\rm{m_{H}/cm^3}$, $r_{\rm{max}}=8\,\rm{kpc}$.
The parameters for the star formation and supernova feedback are $c_\star=0.05$ and $\epsilon_{\rm{SN}}=0.01$.
}
\label{fig12_appendix_2}
\end{figure*}

\begin{figure*}
\resizebox{\hsize}{!}{\includegraphics[angle=0]{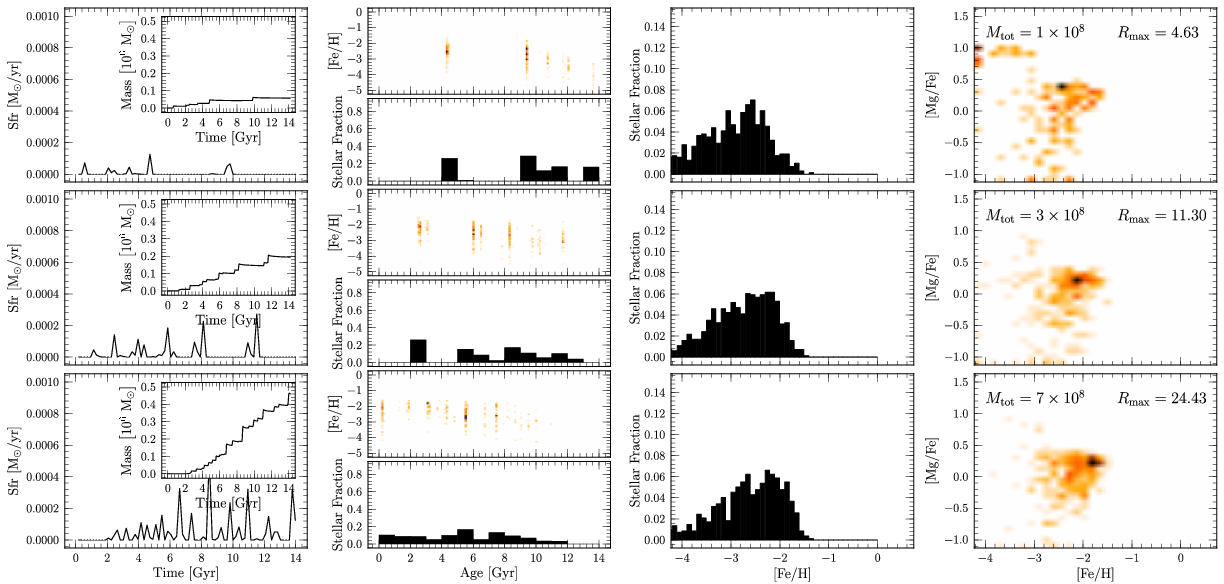}}
\caption{Effect of the total mass. We vary $r_{\rm{max}}$ keeping a constant central gas density $\rho_{\rm{c,gas}}=0.015\,\rm{m_{H}/cm^3}$.
The parameters for the star formation and supernova feedback are $c_\star=0.1$ and $\epsilon_{\rm{SN}}=0.05$.}
\label{fig13_appendix_2}
\end{figure*}

\begin{figure*}
\resizebox{\hsize}{!}{\includegraphics[angle=0]{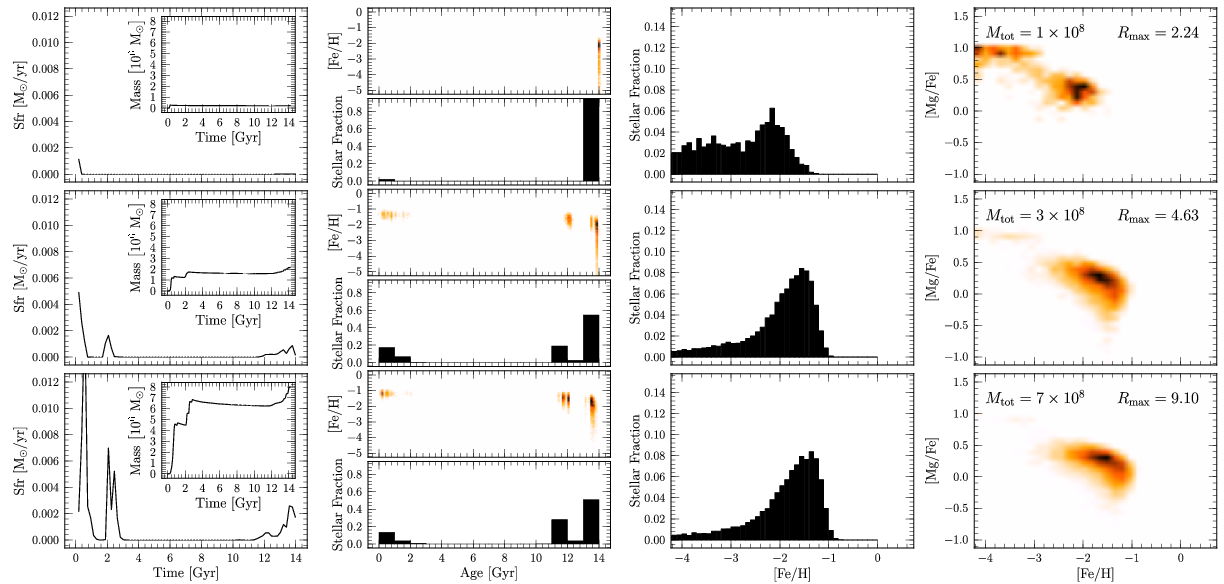}}
\caption{Effect of the total mass. We vary  $r_{\rm{max}}$, keeping a constant central gas density $\rho_{\rm{c,gas}}=0.044\,\rm{m_{H}/cm^3}$.
The parameters for the star formation and supernova feedback are $c_\star=0.1$ and $\epsilon_{\rm{SN}}=0.05$.}
\label{fig15_appendix_2}
\end{figure*}

\begin{figure*}
\resizebox{\hsize}{!}{\includegraphics[angle=0]{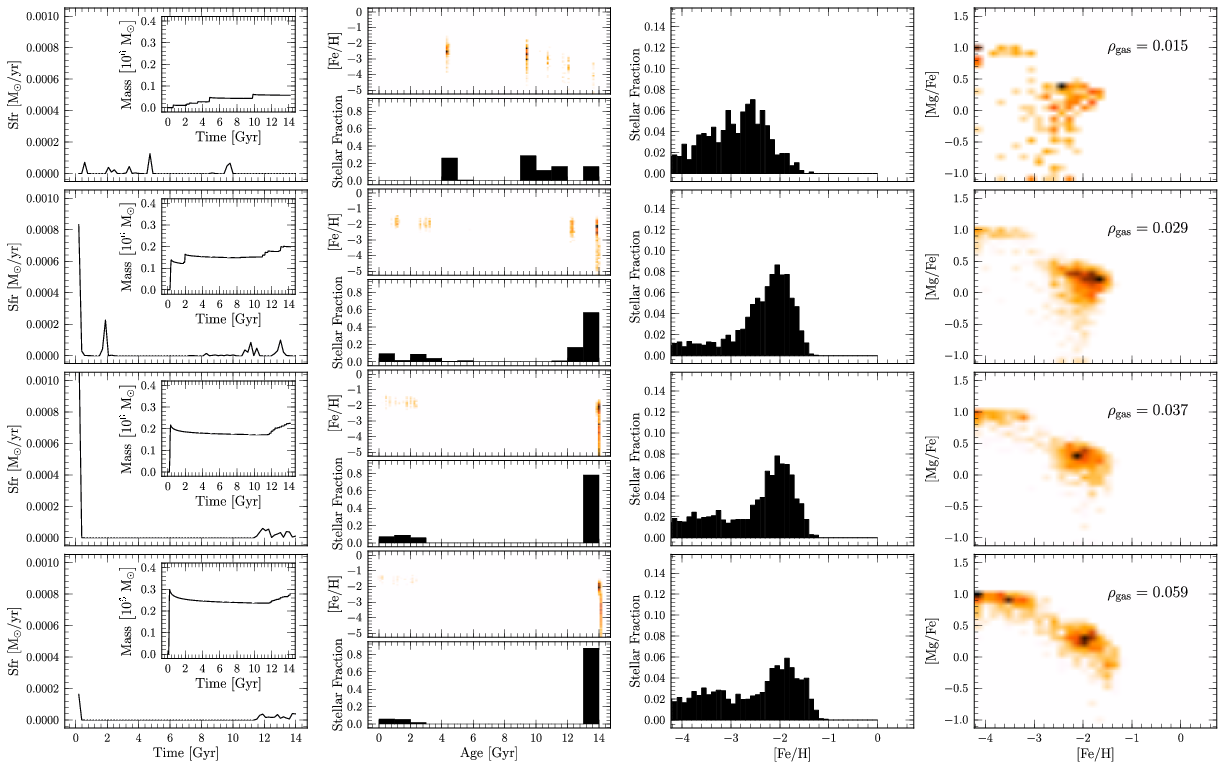}}
\caption{Effect of varying the central gas density, for a constant total mass $M_{\rm{tot}}=1\times 10^8\,\rm{M_\odot}$.
The parameters for the star formation and supernova feedback are $c_\star=0.1$ and $\epsilon_{\rm{SN}}=0.05$.}
\label{fig17_appendix_2}
\end{figure*}

\begin{figure*}
\resizebox{\hsize}{!}{\includegraphics[angle=0]{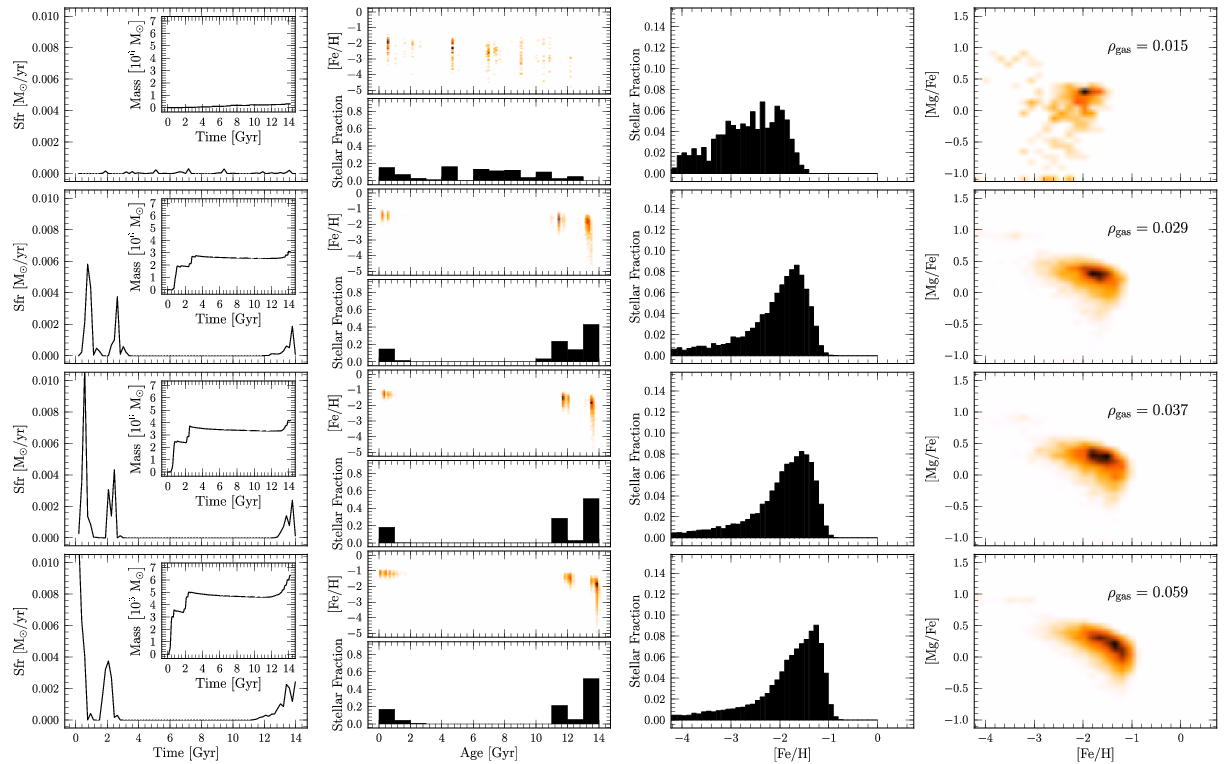}}
\caption{Effect of varying  the central gas density, for a constant total mass $M_{\rm{tot}}=5\times 10^8\,\rm{M_\odot}$.
The parameters for the star formation and supernova feedback are $c_\star=0.1$ and $\epsilon_{\rm{SN}}=0.05$.}
\label{fig19_appendix_2}
\end{figure*}

\end{appendix}

\bibliographystyle{aa}
\bibliography{bibliography}

\end{document}